\documentstyle[epsfig]{mn}
\newif\ifAMStwofonts

\ifoldfss
  \ifCUPmtlplainloaded \else
    \NewTextAlphabet{textbfit} {cmbxti10} {}
    \NewTextAlphabet{textbfss} {cmssbx10} {}
    \NewMathAlphabet{mathbfit} {cmbxti10} {} 
    \NewMathAlphabet{mathbfss} {cmssbx10} {} 
  \fi
  \ifAMStwofonts
    \ifCUPmtlplainloaded \else
      \NewSymbolFont{upmath} {eurm10}
      \NewSymbolFont{AMSa} {msam10}
      \NewMathSymbol{\upi}     {0}{upmath}{19}
      \NewMathSymbol{\umu}     {0}{upmath}{16}
      \NewMathSymbol{\upartial}{0}{upmath}{40}
      \NewMathSymbol{\leqslant}{3}{AMSa}{36}
      \NewMathSymbol{\geqslant}{3}{AMSa}{3E}

       \let\le=\leqslant
       \let\ge=\geqslant
    \fi
  \fi
\fi 

\ifnfssone
  \newmathalphabet{\mathit}
  \addtoversion{normal}{\mathit}{cmr}{m}{it}
  \addtoversion{bold}{\mathit}{cmr}{bx}{it}
  \newmathalphabet{\mathbfit} 
  \addtoversion{normal}{\mathbfit}{cmr}{bx}{it}
  \addtoversion{bold}{\mathbfit}{cmr}{bx}{it}
  \newmathalphabet{\mathbfss} 
  \addtoversion{normal}{\mathbfss}{cmss}{bx}{n}
  \addtoversion{bold}{\mathbfss}{cmss}{bx}{n}
  \ifAMStwofonts
    \ifCUPmtlplainloaded \else
      \UseAMStwoboldmath
      \makeatletter
      \new@mathgroup\upmath@group
      \define@mathgroup\mv@normal\upmath@group{eur}{m}{n}
      \define@mathgroup\mv@bold\upmath@group{eur}{b}{n}
      \edef\UPM{\hexnumber\upmath@group}
      \new@mathgroup\amsa@group
      \define@mathgroup\mv@normal\amsa@group{msa}{m}{n}
      \define@mathgroup\mv@bold\amsa@group{msa}{m}{n}
      \edef\AMSa{\hexnumber\amsa@group}
      \makeatother
      \mathchardef\upi="0\UPM19
      \mathchardef\umu="0\UPM16
      \mathchardef\upartial="0\UPM40
      \mathchardef\leqslant="3\AMSa36
      \mathchardef\geqslant="3\AMSa3E

       \let\le=\leqslant
       \let\ge=\geqslant
    \fi
  \fi
\fi 

\ifnfsstwo
  \DeclareMathAlphabet{\mathbfit}{OT1}{cmr}{bx}{it}
  \SetMathAlphabet\mathbfit{bold}{OT1}{cmr}{bx}{it}
  \DeclareMathAlphabet{\mathbfss}{OT1}{cmss}{bx}{n}
  \SetMathAlphabet\mathbfss{bold}{OT1}{cmss}{bx}{n}
  \ifAMStwofonts
    \ifCUPmtlplainloaded \else
      \DeclareSymbolFont{UPM}{U}{eur}{m}{n}
      \SetSymbolFont{UPM}{bold}{U}{eur}{b}{n}
      \DeclareSymbolFont{AMSa}{U}{msa}{m}{n}
      \DeclareMathSymbol{\upi}{0}{UPM}{"19}
      \DeclareMathSymbol{\umu}{0}{UPM}{"16}
      \DeclareMathSymbol{\upartial}{0}{UPM}{"40}
      \DeclareMathSymbol{\leqslant}{3}{AMSa}{"36}
      \DeclareMathSymbol{\geqslant}{3}{AMSa}{"3E}

       \let\le=\leqslant
       \let\ge=\geqslant
    \fi
  \fi
\fi 

\ifCUPmtlplainloaded \else
  \ifAMStwofonts \else 
    \def\upi{\pi}
    \def\umu{\mu}
    \def\upartial{\partial}
  \fi
\fi

\title{The stellar metallicity distribution of disc galaxies and bulges 
in cosmological simulations} 
\author[F. Calura, et al.  ]
       {F. Calura$^{1,2}$\thanks{E-mail: fcalura@oabo.inaf.it}, 
        B.K. Gibson$^{2,3,4}$, 
        L. Michel-Dansac$^{5}$,
        G.S. Stinson$^{2,6}$,\newauthor
        M. Cignoni$^{1,7}$,
        A. Dotter$^{8}$,
	K. Pilkington$^{2,3,4}$,
        E.L. House$^{2}$,
        C.B. Brook$^{2,9}$,
        C.G. Few$^{2}$,\newauthor
        J. Bailin$^{10}$, 
        H.M.P. Couchman$^{11}$, and
        J. Wadsley$^{11}$ \\
$^{1}$INAF, Osservatorio Astronomico di Bologna, via Ranzani 1, 40127 Bologna, 
Italy\\
$^{2}$Jeremiah Horrocks Institute, University of Central Lancashire, Preston, 
PR1~2HE, UK\\
$^{3}$Department of Astronomy \& Physics, Saint Mary's University, Halifax,
Nova Scotia, B3H~3C3, Canada\\
$^{4}$Monash Centre for Astrophysics, School of Mathematical Sciences, 
Monash University, Clayton, VIC, 3800, Australia\\
$^{5}$Centre de Recherche Astrophysique de Lyon, Universit\'e de Lyon, Obs. 
de Lyon, CNRS, Saint-Genis Laval, 69230, France\\
$^{6}$Max-Planck-Institut f\"ur Astronomie, K\"onigstuhl 17, 69117, Heidelberg, Germany\\ 
$^{7}$Dipartimento di Astronomia, Universit\'a degli Studi di Bologna, via Ranzani 1, 40127 Bologna, Italy\\
$^{8}$Space Telescope Science Institute, 3700 San Martin Drive, Baltimore, MD 21218, USA\\
$^{9}$Departamento de F\'isica Te\'orica, Universidad Aut\'onoma de Madrid,
E-28049 Cantoblanco, Madrid, Spain\\
$^{10}$Astronomy Department, University of Michigan, 500 Church St., 
Ann Arbor, MI, 48109-1042, USA\\
$^{11}$Department of Physics \& Astronomy, McMaster University, 
Hamilton, Ontario, L8S~4M1, Canada}
	
\begin{document}
\date{Submitted}
\pagerange{\pageref{firstpage}--\pageref{lastpage}}
\pubyear{2011}

\maketitle
\label{firstpage}

\begin{abstract}
By means of high-resolution cosmological hydrodynamical simulations of 
Milky Way-like disc galaxies, we conduct an analysis of the associated 
stellar metallicity distribution functions (MDFs). After undertaking a 
kinematic decomposition of each simulation into spheroid and disc 
sub-components, we compare the predicted MDFs to those observed in the 
solar neighbourhood and the Galactic bulge. The effects of the star 
formation density threshold are visible in the star formation histories, 
which show a modulation in their behaviour driven by the threshold. The 
derived MDFs show median metallicities lower by 0.2$-$0.3~dex than the 
MDF observed locally in the disc and in the Galactic bulge. Possible 
reasons for this apparent discrepancy include the use of low stellar 
yields and/or centrally-concentrated 
star formation. The dispersions are larger than the one of the observed MDF; 
this 
could be due to simulated discs being kinematically hotter relative to the Milky Way. 
The fraction of low metallicity stars is largely 
overestimated, visible from the more negatively skewed MDF with respect to the observational sample. 
For 
our fiducial Milky Way analog, we study the metallicity distribution of 
the stars born \it in situ \rm relative to those formed via \it 
accretion \rm (from disrupted satellites), and demonstrate that this 
low-metallicity tail to the MDF is populated primarily by accreted 
stars.  Enhanced supernova and stellar radiation energy feedback to the 
surrounding interstellar media of these pre-disrupted satellites is 
suggested as an important regulator of the MDF skewness.
\end{abstract} 

\begin{keywords}
galaxies: evolution -- galaxies: abundances --- methods: numerical
\end{keywords}

\section{Introduction} 

Galaxy formation and evolution is a distinctly multi-disciplinary field, 
connecting fundamental cosmology and structure formation, through  
stellar astrophysics, nucleosynthesis, and therefore atomic physics.  
These extremes manifest themselves in the appearance and characteristics 
of the stellar populations which comprise the galaxies we observe 
empirically.

Because of the obvious, deep-rooted, interest in understanding the 
origin of our own Milky Way, there are intense efforts underway to 
understand the underlying physics driving disc galaxy formation within 
the concordant $\Lambda$CDM cosmology. From an empirical perspective, 
the Milky Way clearly possesses the greatest wealth of observational 
constraints to any formation scenario, from accurate 6d phase-space 
coordinates (positions and velocities), chemical abundances, and ages, 
for massive numbers of halo, disc, and bulge stars (e.g., Freeman \& 
Bland-Hawthorn 2002).

Galactic chemical evolution models developed in a cosmological framework 
are particularly fruitful tools to derive crucial information regarding 
the star formation histories (SFH) of galaxies, on the ages of the 
stellar populations, and on the gas accretion and outflow histories. A 
key observable for constraining any such model is the metallicity  
distribution function (MDF) of the stars of the various sub-components 
of a galaxy. The MDF bears information concerning the star formation 
history of our Galaxy (e.g. Tinsley 1980; Matteucci \& Brocato 1990; Pagel \& Tautvaisiene 1995; 
Caimmi 1997; Haywood 2006), 
which is directly 
linked to the merging history of its progenitors, i.e. the 
``building blocks'' of the various components.
A detailed study of the impact of the accretion history on the 
shape of the MDF can be found in Font et al. (2006), where it was shown 
that the earlier the major accretion epoch of satellites of 
the central galaxy, 
the more the MDF peak is shifted toward lower metallicities.
These works show how the study of the MDF is important to gain information on 
the evolution of 
both the dark matter content of these systems and of the baryonic 
matter, governed by various physical processes which lead ultimately to 
self-regulated star formation. The metal content of galaxies grows with 
time (modulo dilution effects from metal-poor infall and metal-rich outflows), hence the disc 
MDF allows us to track the enrichment history of the Milky Way 
empirically and the enrichment history of simulated Milky Way-like 
analogs disc through model `deconstruction'.  The extreme, metal-poor 
tail of the MDF provides crucial information on the earliest enrichment 
phases of the disc, while the most metal-rich stars bear the imprint of 
the latest galactic evolutionary stages. 

In what follows, we study the MDFs of the discs and bulges associated 
with a family 
of high-resolution cosmological hydrodynamical simulations. These tools 
represent an ideal instrument to follow the dynamical and chemical 
evolution of the Galactic stellar populations from first principles, 
with a detailed knowledge at any timestep of the spatial distibution of 
gaseous and stellar matter. This facilitates comparison with 
observational data, in particular local solar neighbourhood stars, since 
in the simulations, the position of each stellar particle is known and 
it is hence straightforward to select physical regions whose properties 
can be associated to those observed in the solar neighbourhood. The 
specific simulations employed in our analysis are described in \S2, the 
results presented in \S3, and our conclusions drawn in \S4.

\section{Model Description}

The six simulations employed here are drawn from the MUGS sample 
(Stinson et~al. 2010).  From these, we derive the MDFs associated with 
their analogous `solar neighbourhoods' and `bulges', and contrast them 
with those measured in the Milky Way.  Below, we provide an overview of 
the six simulations, along with the kinematic decomposition employed to 
separate disc stars from spheroid stars.  A more extensive background to 
the simulations is provided by Stinson et~al. (2010), while the radial 
and vertical metallicity gradients are explored by Pilkington et~al. 
(2012a).

\subsection{The MUGS Simulations}

The MUGS simulations were run using the gravitational N-body $+$ 
SPH code \textsc{Gasoline} (Wadsley et~al. 2004). Here, we provide a brief 
overview of the star formation and feedback recipes employed, as
they impact most directly on the chemical abundances associated with
the stellar populations; full details of the simulation are given in 
Stinson et al. (2010). 
The basic star formation and supernova 
feedback follows the `blastwave scenario' of Stinson et~al. (2006);
stars can form from SPH gas particles which meet specific 
density ($>$1~cm$^{-3}$), temperature ($<$15000~K) and convergent flow  criteria.
When these are met, stars can form with a star formation
rate $dM_{\star}/dt$ given by
\begin{equation}
\frac{dM_{\star}}{dt}=c^{\star}\frac{M_{gas}}{t_{dyn}}, 
\end{equation}
where c$^{\star}=0.015$ represents the star formation efficiency. This 
quantity is tuned in order to match the Kennicutt law for star formation 
in local disc galaxies (Kennicutt 1998). The quantity $M_{gas}$ is the 
mass of the star-forming gas particle, while $ t_{dyn}$ is its dynamical 
timescale.

Within each `star' particle, `individual' star masses are distributed
according to a Kroupa, Tout \& Gilmore (1993) initial mass
function (IMF), with lower and upper mass limits of 0.1~M$_\odot$ and
100~M$_\odot$, respectively.  Stars with masses between 8~M$_\odot$ and
40~M$_\odot$ are assumed to explode as Type~II supernovae (SNeII).
Each supernova is assumed to possess an energy of 10$^{51}$~erg, 
and we assume 40\% of this energy is 
made available in the form of thermal energy to the surrounding
interstellar medium (ISM).

The heavy elements restored to the ISM by SNeII in this version of 
\textsc{Gasoline} are O and Fe.  Analytical power-law fits in mass were 
made using the yields of Woosley \& Weaver (1995), convolved with the 
aforementioned Kroupa et~al. (1993) IMF, to derive the mass 
fraction of metals ejected by SNeII for each stellar particle.  These 
elements are returned on the timescale of the lifetimes of the 
individual stars comprising the IMF, after Raiteri et~al. (1996).  
Type~Ia supernovae (SNeIa) are included within \textsc{Gasoline}, 
patterned again after the Raiteri et~al. implementation of the Greggio 
\& Renzini (1983) single-degenerate progenitor model.  Each SNeIa is 
assumed to return 0.63~M$_{\odot}$ of iron and 0.13~M$_{\odot}$ of 
oxygen to the ISM.  
This is an important feature of the code, which differentiates it from 
other previous attempts to model chemical abundances in simulations. 
In the past,cosmological codes tracked the total gas metallicity (Z), 
under the assumption of the instantaneous recycling approximation, 
i.e. neglecting the time delay between star formation and the energetic and 
chemical feedback from stellar winds and SNe (e. g., Sanchez-Blazquez et al. 
2009). Such codes are not suited to study elements such as Fe, 
mainly produced by type Ia SNe on timescales varing from 0.03 Gyr up to several Gyr, but which are crucial since they are primary metallicity tracers, 
in particular in observational studies of the stellar MDF. 
In this paper, we take into account finite time delays for the main 
channels for Fe production, i.e. type Ia and type II SNe, hence our 
study should be regarded as a significant step forward with respect to 
previous studies of chemical abundances in cosmological simulations. 
Other recent chemical evolution studies 
in fully cosmological 
disc simulations relaxing the instantaneous recycling approximation 
include Rahimi et al. (2010) and Few et al. (2012).

The contribution of single low and intermediate 
mass asymptotic giant branch stars is not included in these runs, but 
for the analysis of oxygen and iron, this is negligible with respect to 
the contributions of SNeII and SNeIa.

The total metallicity in this version of \textsc{Gasoline} is tracked by 
assuming Z$\equiv$O+Fe.\footnote{By assuming Z$\equiv$O+Fe, we 
underestimate the global metal production rate by 
roughly a factor of two, which leads to a parallel underestimate in the gas cooling rate, and hence star formation rate 
(Pilkington et al. 2012a). 
Given the strong non-linearity of the dependence of the feedback and cooling 
processes on the global metallicity, the only way to quantify how this alters 
our results would be to re-run all the simulations with the correct chemical evolution 
prescriptions;} to this purpose, our next generation of runs with 
\textsc{Gasoline} will employ a more complete chemical `network', 
ensuring that $\sim$90\% of the global metallicity `Z' will be tracked 
element-by-element (see Pilkington et al. 2012b).
For these runs, only the solar metallicity 
yields were employed, and long-lived SNeIa progenitors (i.e. those in binary systems 
with companions having mass $m$$<$1.5~M$_\odot$) were neglected. For a Kroupa 
et~al. (1993) IMF, in a simple stellar population the amount of Fe 
produced by the progenitors with mass $m$$<$1.5~M$_\odot$ is only 
$\sim$20\% of the amount produced by all progenitors, i.e. with masses 
ranging from $0.8 M_{\odot}$ to $8 M_{\odot}$. While not important for 
our MDF work here, the systematic neglect of long-lived SNeIa 
progenitors \it could \rm have a potential impact on the [O/Fe]-[Fe/H] 
relationship (Pilkington et~al., in preparation).

For our analysis,  
we selected six galaxies with the most prominent discs, following the same criteria as described in Pilkington et al. (2012a)\footnote{The selected galaxies 
are those for which there was unequivocal identification of the disc (from angular momentum
arguments constructed from the gas and young star distributions, see Stinson et al. 2010). 
In this way, we are able to eliminate extreme values
of bulge-to-total, but formally, we only included those disks for which alignment based upon the 
gas/young stars was obvious.}. 
Each of the six MUGS simulations analysed here were run in a 
50$h^{-1}$~Mpc cosmological box with `volume renormalization' to ensure 
higher space and force resolution in the region centred on the central 
galaxy (Klypin et~al. 2001). For each simulation, the $z$=0 output was 
examined to find sufficiently isolated halos within the mass range 
5$\times$10$^{11}$~M$_{\odot}$ and 2$\times$10$^{12}$~M$_{\odot}$. 
Sixteen  halos within this range were selected at random and re-run 
at higher resolution (9 of which are described by Stinson et~al. 2010, 
with a further 7 now having been realised subsequent to the publication 
of the first sample). Six galaxies were chosen from the MUGS suite': 
\tt g1536\rm, \tt g24334\rm, \tt g28547\rm, \tt g422\rm, \tt g8893\rm, 
and \tt g15784\rm.
The system \tt g15784 \rm is our adopted fiducial Milky Way analog, 
owing to its total mass and its baryonic mass in the disc, both similar 
to the values calculated with up-to-date dynamical models of our Galaxy 
(see McMillan 2011). 

In Table~\ref{tab_mugs}, we list the key properties for each of the simulations 
employed here. Following the same notation as the original MUGS 
simulations, galaxies are identified using the group number from the 
original friends-of-friends galaxy catalogue. The first column contains 
the galaxy name; the second, third, and fourth columns are, for each 
galaxy, the total (baryonic and non-baryonic) mass, the gas mass, and 
the stellar mass inside the virial radius $R_{vir}$, respectively. The 
fifth, sixth, and seventh columns are the corresponding number of dark 
matter particles, gas particles, and stellar particles, respectively. 
The eigth and ninth columns are the total disc mass and the total bulge mass 
assigned to each sub-component, after application of the kinematic 
decomposition and spatial cuts described in Sections~2.2 and 2.3.
The final two columns are the half mass radius for the spheroid component ($R_{eff}$) and the disc scalelength ($R_d$), 
calculated by means of an exponential profile fit to the disc component.

\subsection{Kinematic Decomposition}
\label{kcut}

To isolate the stellar components associated with the simulated bulge 
and disc for each simulation, we performed kinematic decompositions, 
after Abadi et~al. (2003).  We first centre and align the angular 
momentum vectors of the baryons with the $z$-axis of the volume, and
remove any systemic velocity associated with the simulated galaxy, for 
ease of subsequent decomposition.

\begin{figure*}
  \centering \includegraphics[width=168mm]{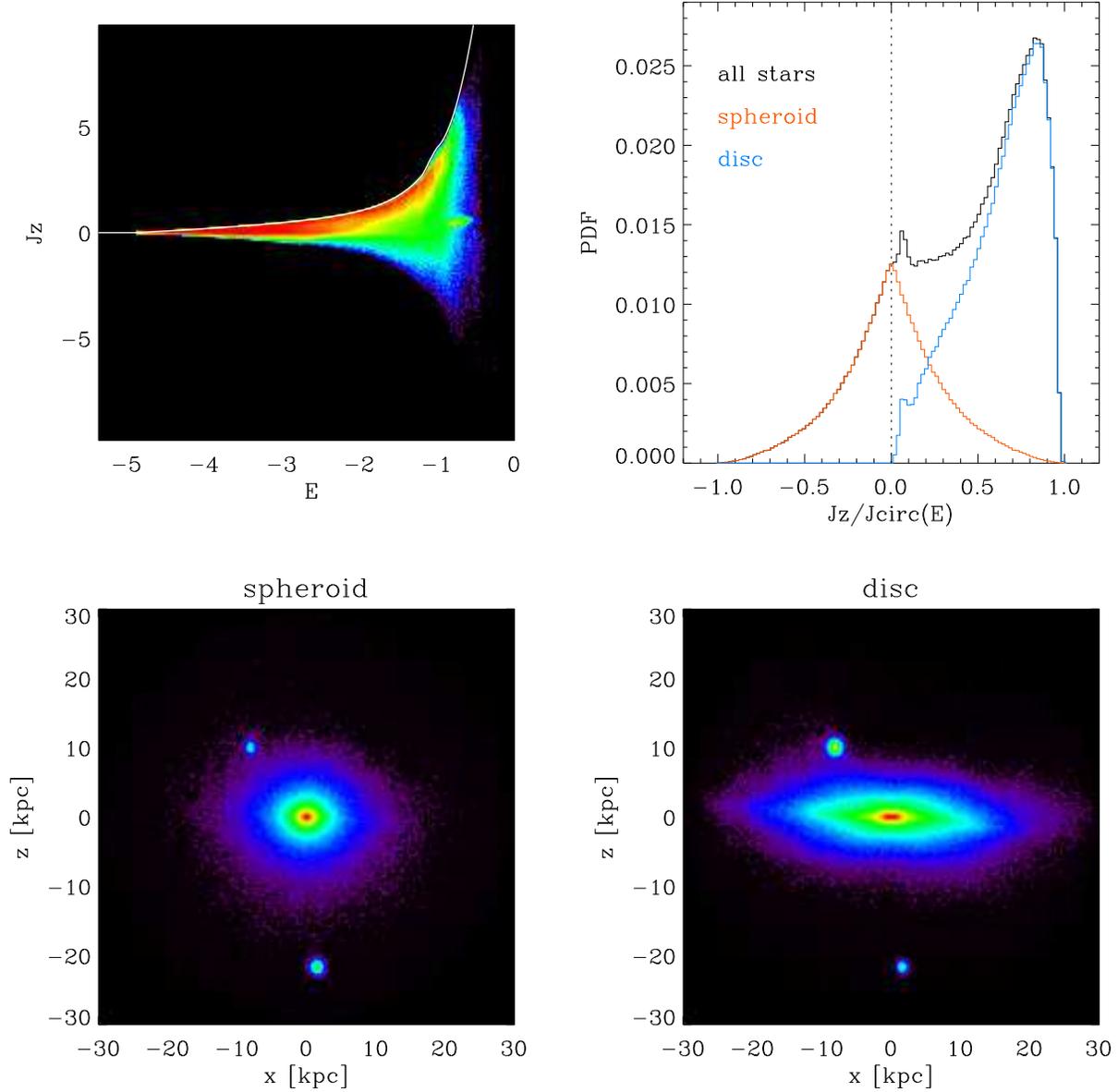}
  \caption{Kinematic decomposition of \tt g15784\rm. Top left panel: the
    $z$-component of the specific angular momentum shown as a function
    of specific binding energy for all stars within 30~kpc of the
    centre of the galaxy. The white curve shows the $J_{circ}(E)$
    location. The colours scale with the density of particles in each
    pixel, with redder colours corresponding to higher densities. Top right
    panel: distribution of the orbital circularity of the stellar
    component (black line) and the decomposition into spheroid (orange
    line) and disc (blue line) components. Bottom panels: images of
    the spheroid (left) and disc (right) components. The colours scale
    with the density of particles in each pixel, with redder colours
    corresponding to higher densities. The sub-structure/peak near $J_{z}/J_{circ}(E) \sim+0.1$ is 
    associated with the satellite seen at $(x,z) \sim(-10,+10)$ in the lower-right panel.}
  \label{decomp}
\end{figure*}

We next compute the Lindblad Diagram of all stellar particles in the 
inner region of the halo ($r$$<$30~kpc), i.e. the $z$-component of the 
specific angular momentum as a function of the specific binding energy. 
An example is shown in Fig~\ref{decomp} (top left panel) for the 
fiducial simulation \tt g15784\rm.  The distribution of the orbital 
circularity $\epsilon_J$ is then constructed, where $\epsilon_J = J_z / 
J_{circ}(E)$ where $J_z$ is the $z$-component of the specific angular 
momentum and $J_{circ}(E)$ the angular momentum of a circular orbit at a 
given specific binding energy. The $J_{circ}(E)$ curve (shown as a white 
line in the top-left panel of Fig.~\ref{decomp}) indicates the location 
of circular orbits corotating with the disc.

The spheroid component (comprised of both bulge and inner halo stars) is 
defined \it a priori \rm to be a symmetric distribution centered on 
$J_z/J_{circ}(E)=0$.  This means that, by construction, the 
bulge/inner halo components are assumed to be non-rotating. Since we are 
here interested in distinguishing the bulge from the disc component, we 
assume that the disc is made by all the stars except those belonging to 
the bulge. Star particles with negative circularity are assigned to the 
spheroid component; those with positive circularity are randomly 
assigned to either the spheroid or disc components, weighted by the 
likelihood imposed by the relative numbers of both components at a given 
positive $J_z/J_{circ}(E)$. To prevent stars of nearby satellites 
from being included in our analysis, we perform a further spatial cut as 
described in ~\ref{spcut}.

This method, while somewhat arbitrary in its definition of the 
components, does allow one to decompose in an objective, 'hands-off' manner, 
the stellar component 
into spheroidal and disc component, as shown graphically in the bottom 
panels of Fig.~\ref{decomp}.

\subsection{Spatial Cuts}
\label{spcut}

In addition to the kinematic decomposition of Sect.\ref{kcut}, for each 
galaxy we also applied a spatial cut to both the spheroid and disc 
components. Spheroid stars within $R_{b,cut}$=1$-$3~kpc are assigned to 
the `bulge', a radial `cut' which qualitatively corresponds to the 
spatial extent of the simulated bulge, although we have confirmed that 
the exact value selected within this range does not impact on our 
conclusions. Further, for each simulation, we assign stars to the disc 
should they lie within 4~kpc of the mid-plane, beyond the aforementioned 
bulge-disc radial cut, i.e. $r' = \sqrt{X^2+Y^2}>R_{b,cut}$.

For one simulation (\tt g24334\rm), an additional spatial cut was 
applied, in order to remove the presence of a dwarf satellite which, at 
redshift $z$=0, is passing through the disk at $r^\prime$$\sim$5~kpc. 
The disc for \tt g24334 \rm was, instead, defined by 
$R_{b,cut}$$<$$r^\prime$$<$4~kpc.  

\begin{table*}
\vspace{0cm}
\begin{flushleft}
\caption[]{Main features of the simulated galaxies from the MUGS sample. All masses are expressed in units of $10^{10} M_{\odot}$ and all radii are in kpc. }
\begin{tabular}{l|lllllll|ll|lll}
\noalign{\smallskip}
\hline
\hline
\noalign{\smallskip}
 Galaxy             &   Total Mass           &  $M_{gas}$         & $M_{*}$            & $N_{DM}$       & $N_{gas}$   &    $N_{*}$   &        & $M_{*,d}$     &  $M_{*,b}$ & & $R_{eff}$ &  $R_{d}$       \\   
                    &  ($10^{10} M_{\odot}$)   &   $(R\le r_{vir})$ &  $(R\le r_{vir})$   & $(10^5)$      & $(10^5)$   &   $(10^5)$   &        &             &            & &  (kpc)    &   (kpc)  \\       
\noalign{\smallskip}                                                                                                                                         
\noalign{\smallskip}                                                                                                                                         
\hline                                                                                                                                                       
\noalign{\smallskip}                                                                                                                                         
\tt g1536  \rm              &    70                  &     5.1         &  6.0          &   5.3            &   2.4        &   13.6        &    &    3.2       &   2.5    &   &   1.3    &     2.5    \\
\tt g15784 \rm              &    140                 &     10          &   11          &   10.8           &   4.8        &   26          &   &    5.9       &   3.1    &   &   1.3    &     3.2     \\ 
\tt g24334 \rm              &    108                 &     7.1         &   10.4        &   8.2            &   3.3        &   24          &   &    1.1       &   1.2    &   &   1.6    &     1.0     \\
\tt g28547 \rm              &    107                 &     7.4         &  11           &   8.1            &   3.4        &   26          &   &    1.1       &   1.3    &   &   1.1    &     2.9     \\ 
\tt g422   \rm              &    91                  &     7.0         &  8.4          &   6.9            &   3.2        &   19.2        &   &    0.9       &   1.9    &   &   2.0    &     2.8     \\
\tt g8893  \rm              &    61                  &     4.3         &  6.1          &   4.6            &   1.9        &   14          &    &    1.2       &   1.3    &   &  1.3     &    2.9     \\
\hline
\hline
\end{tabular}
\label{tab_mugs}
\end{flushleft}
\end{table*}

\begin{figure*}
\epsfig{file=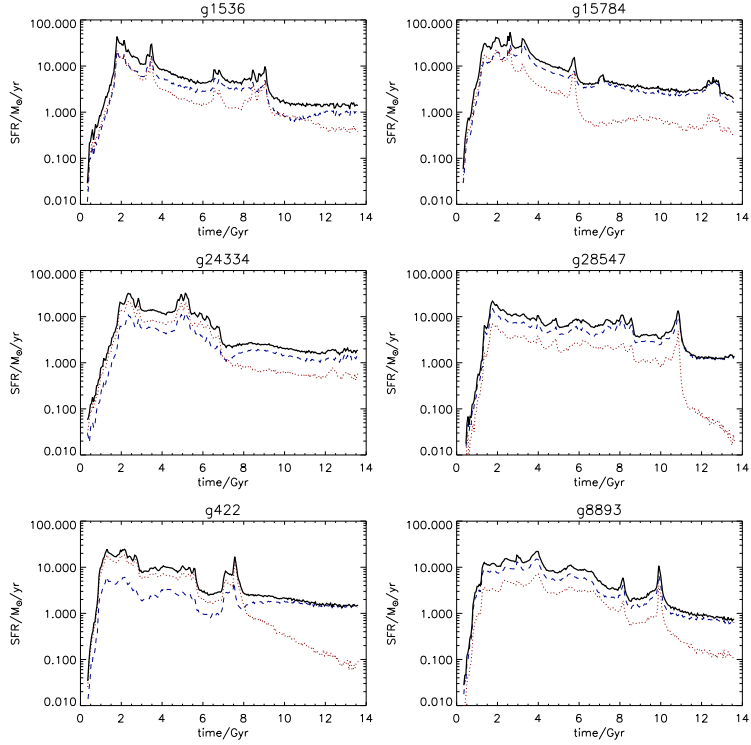,height=10cm,width=10cm}
\caption{Star formation histories (SFHs) of the MUGS galaxies considered 
in this work. In each panel, the black solid lines are SFHs computed for all the stars inside 
$R_{vir}$. The blue dashed lines and the red dotted lines are 
the SFHs for stars belonging to the disc and the bulge, respectively,  
after performing a kinematic cut as described in Sect.~\ref{kcut}.}
\label{sfr}
\end{figure*}
\begin{figure*}
\epsfig{file=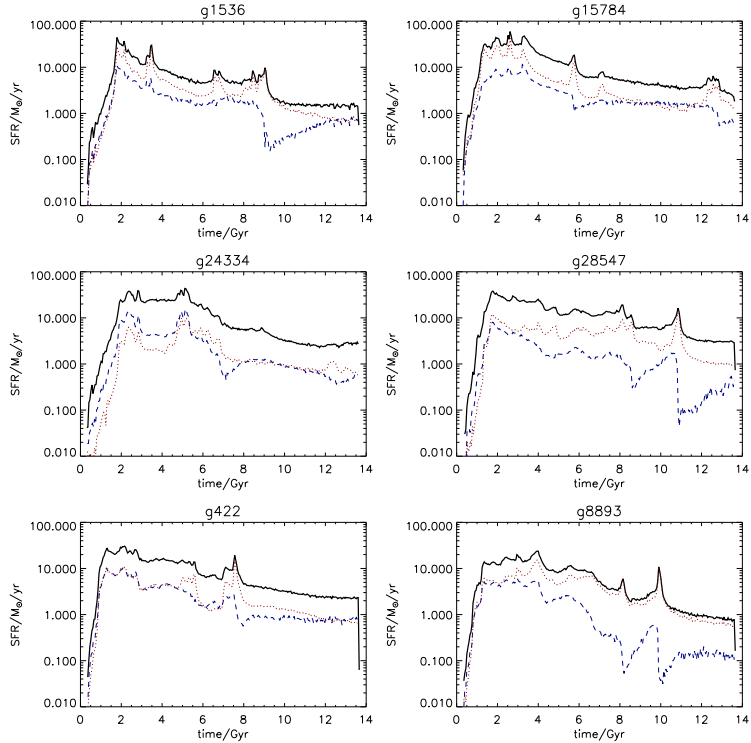,height=10cm,width=10cm}
\caption{Star formation histories (SFHs) of the MUGS galaxies considered 
in this work. In each panel, the black solid lines are 
as in Fig. ~\ref{sfr}, 
whereas the blue dashed lines and the red dotted lines are 
the SFHs for stars belonging to the disc and the bulge, respectively,  
after performing a pure spatial cut as described in Sect.~\ref{spcut}.}
\label{sfr_puresp}
\end{figure*}
\begin{figure*}
\epsfig{file=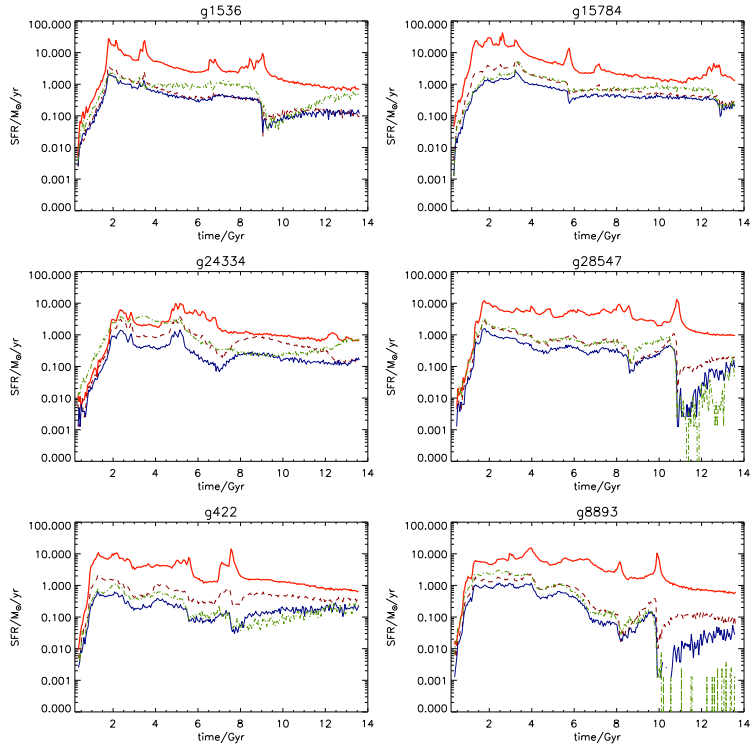,height=10cm,width=10cm}
\caption{Star formation histories (SFHs) of the MUGS discs in different 
spatial regions. 
The red dashed lines, blue solid lines, and green dash-dotted 
lines represent the SFHs computed considering the star particles which 
at the present time are located at radii $r$$<$2~$R_{d}$, between 
2~$R_{d}$ and 3~$R_{d}$, and radii $r$$>$3~$R_{d}$, respectively. 
To trace the SFH of the innermost disc regions, disc particles within typical 
bulge radii (i.e. at $r$$<$1-3 kpc) have been excluded.  
The thick solid red lines are pure spatially cut bulge SFHs, i.e. they represent SFHs 
calculated considering all the star particles in the innermost ``bulge'' regions, regardless of their 
kinematics.}
\label{sfr_spat}
\end{figure*}
\begin{figure*}
\epsfig{file=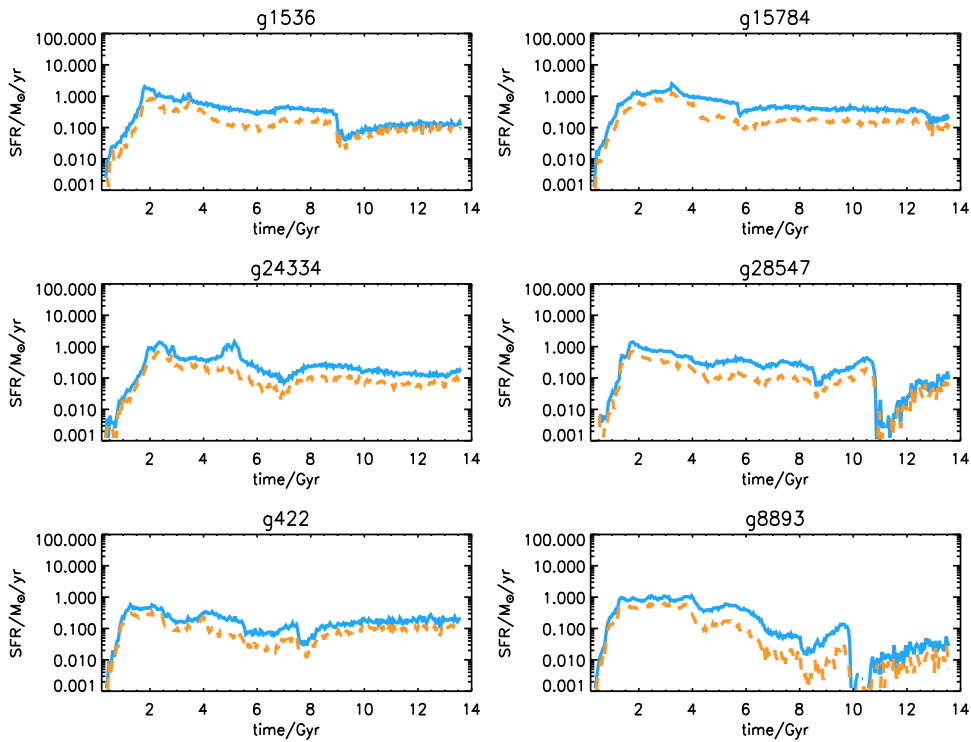,height=10cm,width=13cm}
\caption{``Solar neighbourhood'' star formation histories of the MUGS galaxies and effect of 
bulge stars with positive circularities. The solid lines are the SFHs calculated after a kinematic 
cut as described in Sect.~\ref{kcut} plus a spatial cut as in Sect.~\ref{spcut}. 
The dashed lines are the result of a spatial cut and the exclusion of all the star particles 
with $J_{z}/J_{circ}(E) < 0.8$. }
\label{sfr_jjc}
\end{figure*}

\section{Results}
\subsection{Star Formation Histories}

In Fig.~\ref{sfr}, the star formation histories (SFHs) of the six MUGS 
galaxies are shown. In each panel, we show the SFH computed considering 
all the star particles within the virial radius (solid lines), that of 
the disc (dashed lines), and the bulge (dotted lines).  
The disc and bulge SFHs reported in Fig.~\ref{sfr} have been 
derived after performing the kinematic decompostion described in Sect.~\ref{kcut}.

All the SFHs 
show similar behaviour, i.e. an $\sim$20$-$30~M$_\odot$/yr peak during 
the first few Gyrs, followed by an exponential and fairly continuous 
decline at later times with a timescale ranging from $\sim$4$-$7~Gyr.

In two cases (\tt g24334 \rm and \tt g422\rm), the early evolutionary 
phases are dominated by more intense centrally-concentrated star 
formation. In the other systems, throughout their whole history (in 
general), the discs show SFRs higher than those of their associated 
bulges.  Moreover, the disc SFHs show higher present-day SFRs than the 
bulges by factors of a few to ten. In a few cases, such as \tt g1536\rm, 
the bulge SFHs are somewhat higher than those encountered in nature, at 
least over the last $\sim$5~Gyrs of the simulation. 
In general, the present-day bulge SFRs are 
higher than those seen in the sample of Fisher et al (2009) by factors of a few, 
except for g28547, which sits at the median of the aforementioned sample.  
That said, it is 
important to remind the reader that in this suite of simulations, the 
only form for star formation `quenching' present is that associated with 
feedback from SNe; in general, this is not sufficient for producing 
passive spheroids in cosmological simulations (e.g. Kawata \& Gibson 
2005) or semi-analytic models (e.g. Calura \& Menci 2009).

In Fig.~\ref{sfr_puresp}, 
we show the SFHs for discs and bulges 
obtained solely via the spatial cut criteria described
in Sect. ~\ref{spcut}. 
All the MUGS galaxies suffer an excessive central concentration of mass and of 
star formation, as already reported by Stinson et al. (2010), who performed 
a careful analysis of the rotation curves of various systems, and of Pilkington et al. (2012a), 
who studied the radial metallicity gradients.

As Fig.~\ref{sfr_puresp} shows, if we assume that all the star particles included within the innermost 1-3 kpc 
belong to the bulges, we end up with unnaturally high present-day 
values for the bulge SFHs, which reflect this well-known issue of 
an excess of mass in the centre of the simulated galaxies. 
Several works have already shown that the problem regarding 
the central concentration of 
mass may be partially alleviated by increasing the 
simulation resolution (e.g. Pilkington et al. 2011; Brook et al. 2012). 
Further tests are needed to understand to what extent resolution may help in ameliorating this 
problem. 

As reported by Stinson et al. (2010), in general MUGS 
bulges are bluer than real bulges, and that 
quenching star formation sooner would produce bulges 
that better match the red sequence. 
In fact, 
MUGS galaxies do not include AGN feedback,
which might significantly help driving star-forming gas out
of the central regions of galaxies and limiting the mass of bulges, as already shown in semi 
 analytic models 
and even on mass scales comparable to the one of the MW bulge (e.g., Calura \& Menci 2011). 

In Fig.~\ref{sfr_spat}, we show the SFHs for different spatial regions 
within each of the galaxies in our sample. 
For each stellar particle, we 
have calculated its present-day distance from the centre of the disc and 
divided the disc into three different regions: (1) the innermost one, in 
which all the particles lie within the inner 2~$R_{d}$, 
excluding those in disc regions within radii typical of bulges (1-3 kpc); 
(2) an annulus 
encompassing the particles with distances 
2~$R_{d}$$<$$r$$<$3~$R_{d}$, and (3) the disc outskirts, including 
all the particles with distances beyond 3~$R_{d}$.

In each panel of 
Fig.~\ref{sfr_spat}, 
the thick solid red lines represent pure spatially-cut bulge SFHs, i.e. SFHs 
calculated considering all the star particles in the innermost ``bulge'' regions, regardless of their 
kinematics.
It is important to note that in Fig.~\ref{sfr_spat} we are showing  
SFHs calculated in different spatial regions considering the present-day 
position of each stellar particle, and not the position where the stellar particles were formed. 
The bulge SFHs dominate the 
overall star formation budget througout most of the cosmic time. 

After having excluded the star particles in the bulge regions, 
the SFHs of the innermost disc regions are in general comparable to those at larger distances.
At late times, the outermost parts tend to show 
`oscillating' SFHs at late times ($>$5~Gyrs). This effect is due to the 
adoption of a star formation density threshold ($>$1~cm$^{-3}$); the 
outermost regions of the discs are characterised by lower densities, 
hence more likely to present such a modulating star formation behaviour, 
once the gas density is comparable to that of the threshold value. This 
phenomenon can also be seen in non-cosmological chemical evolution 
models for disc galaxies (e.g. Chiappini et~al. 2001; Cescutti et~al. 
2007) which adopt a density threshold for star formation.  We refer the
reader to a more targeted analysis of the temporal evolution of the 
radial star formation rate profiles within simulated cosmological
discs by Pilkington et~al. (2012a).
It is also important to recall that radial migration is likely to occur 
within these simulations.  Sanchez-Blazquez et~al. (2009) presented the 
analysis of a cosmological disc simulation, comparable in mass and 
kinematic heating profile (e.g. House et~al. 2011) to those analysed 
here, and found that the mean radial distance traversed by the disc star 
particles was $\sim$1.7~kpc. The effects of stellar migration are not taken into account 
in Fig.~\ref{sfr_spat}. 

To assess the effect of bulge stars with 
positive circularities, in Fig. ~\ref{sfr_jjc} we show the ``solar neigbourhood'' SFHs 
computed as in Fig.~\ref{sfr_spat}, compared to the SFHs computed considering 
the star particles 
 with 2~$R_{d}$$<$$r$$<$3~$R_{d}$ and having excluded the ones with 
$J_z/J_{circ}(E)<0.8$. As visible in Fig. 1, at this $J_z/J_{circ}(E)$ value the 
bulge circularity distribution is very low, whereas the one for the disc is close to 
the peak value. 
This ensures the removal of a substantial fraction of particles with bulge kinematics, 
while at the same time, retaining the majority of disc particles.

The difference bewteen the two SFHs is in most of the cases more visible at early (i.e. $<$ 8 Gyr) 
times, when the bulge SFH was particularly intense, as seen in Fig. ~\ref{sfr_spat}. 
Overall, the exclusion of star particles with $J_z/J_{circ}(E)<0.8$ does not seem to affect the 
global shape of the SFHs. 

In the current analysis, the use 
of a spatial or kinematical definition of the 
solar neighbourhood region does not substantially affect our results on the metallicity distribution. 
There is a low contamination from low angular momentum, high metallicity bulge star  
particles at large distances from the centre. 
The effects of bulge stars with positive circularities on the solar neighbourhod MDF 
will be studied in detail later in Sect.~\ref{SN}.

\subsection{Metallicity Distribution Functions in the Discs}

\begin{figure*}
\epsfig{file=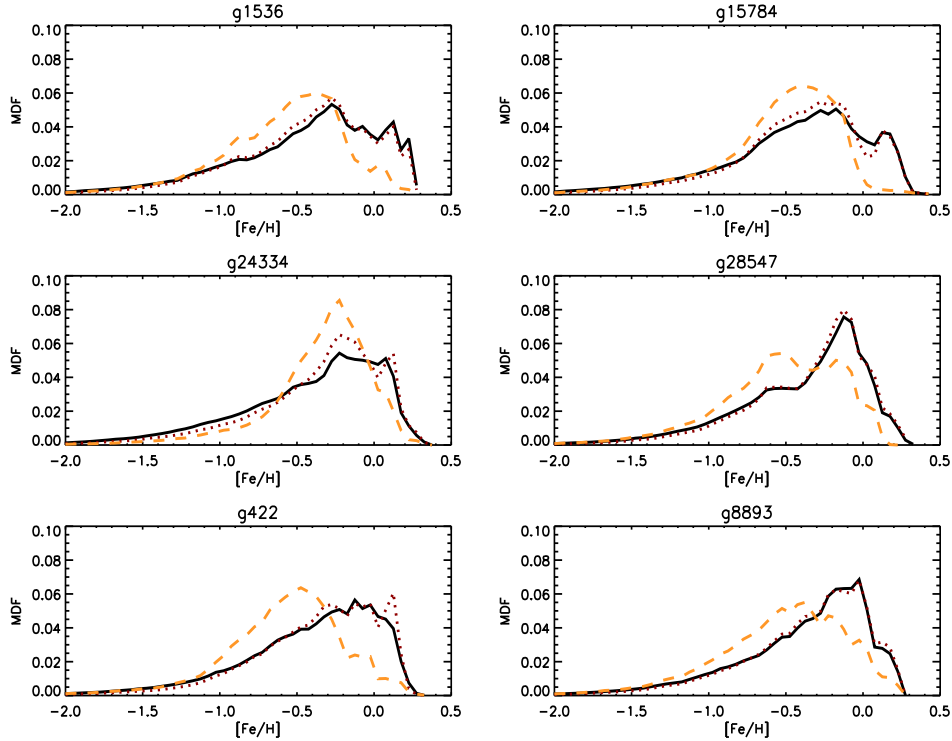,height=10cm,width=13cm}
\caption{Metallicity distribution functions for the six MUGS discs. 
Thick solid lines: MDF calculated considering all the stellar particles 
within the virial radius. Thin dotted lines: MDFs for the particles 
belonging to the discs after the kinematic decomposition described in 
\S\ref{kcut}. Thin dashed lines: MDF calculated for particles in the 
disc and after the additional spatial cut described in \S\ref{spcut}, 
i.e. considering the star particles in the disc within 4~kpc of the 
mid-plane and at galactocentric radii R$\ge R_b$.}
\label{mdf}
\end{figure*}

In Fig.~\ref{mdf}, we show the MDFs of our six simulated galaxies. In 
each panel, we show the MDF calculated using all the stellar particles (i) 
located within the virial radius $R_{vir}$, (ii) in 
the disc after the kinematic decomposition described in Sect.
~\ref{kcut} and (iii) after a kinematic descomposition plus a spatial cut as described in Sect. ~\ref{spcut}.  

The MDFs for the particles included within $R_{vir}$ show several peaks, 
each corresponding to stellar populations associated with various 
kinematic components.  A representative case is that of our fiducial 
simulation \tt g15784\rm, whose MDF shows a high-metallicity peak at 
[Fe/H]$\sim$0.2 and a broader, more significant, peak at lower 
metallicity (near [Fe/H]$\sim$$-$0.2). The disc MDFs are in most cases very similar to the ones calculated 
using all the star particles within $R_{vir}$. 

Performing the spatial cut of \S\ref{spcut}, we can see that the 
high-metallicity peak of the MDF has been removed (\tt g1536 \rm and \tt 
g15784\rm) or substantially reduced (\tt g28547\rm). This is because  
in any galaxy, the highest metallicity stellar particles tend to reside 
near the centre, similar indeed to what is encountered in nature, 
including our own Milky Way, where the metal-rich stars are found 
preferentially in the bulge and inner disc.  

In some cases (\tt g28547 \rm and \tt g8893\rm), the multi-peaked 
structure of the MDF is still present (or even exacerbated) after 
performing the spatial cut. This is due to their particular SFHs, which 
tend to show several late-time star formation episodes. A similar behaviour 
was found for dwarf galaxies within a semi-analytic galaxy formation 
model (Calura \& Menci 2009), where multi-peak SFHs tend to be 
associated with complex multi-component stellar metallicity 
distributions.

Finally, we stress that a comprehensive analysis of 
the origin of the metallicity gradients in the MUGS discs, 
including a few cases described here, can be found in Pilkington 
et~al. (2012a). 
In general, MUGS galaxies can account for the slope of the 
metallicity gradient observed today in young stars in the Milky Way and in HII regions 
in local discs. The analysis of Pilkington et al. (2012a) showed that the 
metallicity zero-point of the MUGS galaxies is offset by 0.2-0.3 dex from those in nature, 
but this does not impact on the determination of the gradients therein.

\subsection{Metallicity Distribution Functions in the Solar Neighbourhood}
\label{SN}
\begin{figure*}
\epsfig{file=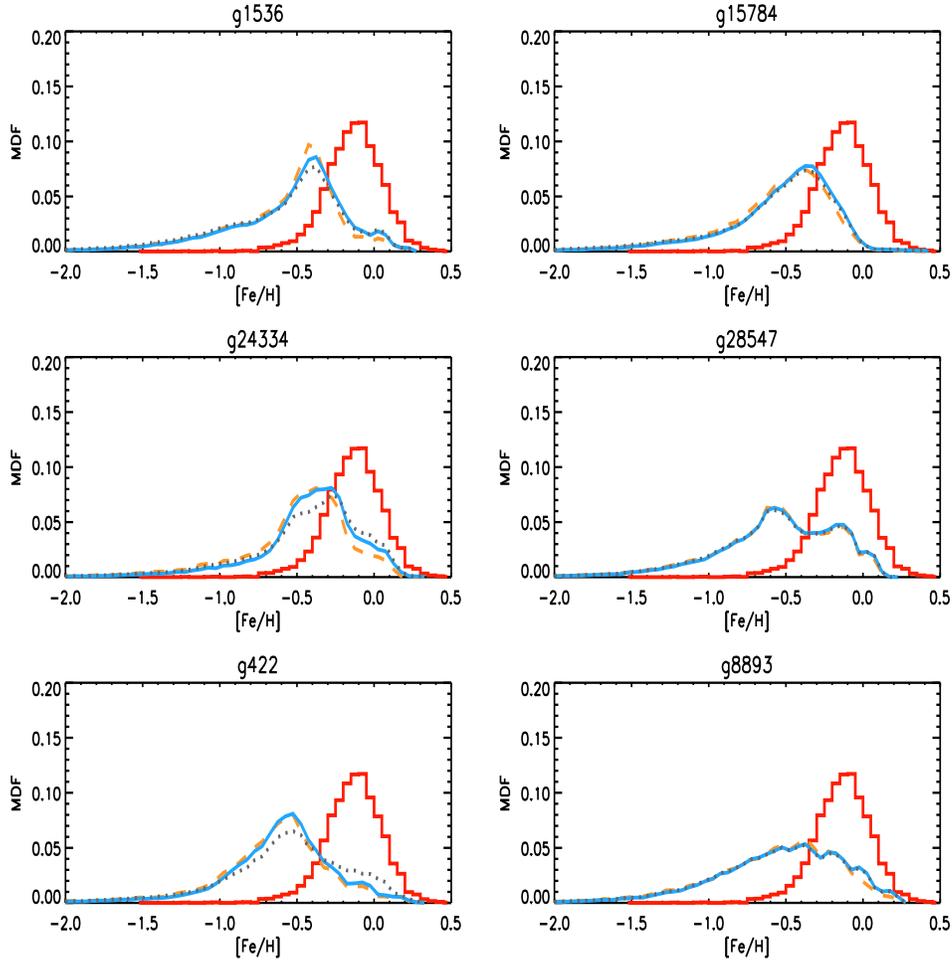,height=13cm,width=13cm}
\caption{Metallicity distribution functions for the 'solar neighbourhood'
of each of the six MUGS discs, compared to the MDF 
observed in the solar neighbourhood as for 
the 'clean' sub-sample drawn  
from the 
Geneva-Copenhagen Survey (Holmberg et al. 2009). 
The solid histrograms represent the observed MDF. 
The solid lines are MDFs calculated after a kinematic 
cut as described in Sect.~\ref{kcut} plus a spatial cut (i.e. 
considering an annulus 
encompassing all the particles with distances $r$ from the centre in the range 
2~$R_{d}$$<$$r$$<$3~$R_{d}$) 
as in Sect.~\ref{spcut}. 
The Dashed lines are MDFs after a spatial cut and the exclusion of all the star particles 
with $J_{z}/J_{circ}(E) < 0.8$. 
Dotted lines: MDFs obtained with a solely spatial cut, i.e. considering the star particles in the annulus without 
any other kinematical selection. }
\label{mdf_sn}
\end{figure*}

In Fig.~\ref{mdf_sn}, we show the MDFs calculated for each simulation 
using star particles situated within a circular annulus of 
2$R_{d}$$<$$r$$<$3$R_{d}$, compared with the observational MDF in the solar neighbouhood. 
The observational MDF adopted here is calculated from the 
Geneva Copenhagen survey (GCS)
sample of solar-neighbourhood stars (Nordstr\"om et al.
2004; Holmberg, Nordstr\"om \& Andersen 2007). 
Nordstr\"om et al. (2004) obtained
Str\"omgren photometry and radial velocities for a magnitude-limited
sample of $\sim$17000 F and G dwarfs.
From the photometry, they estimated metallicities
and ages. There has been some debate about the calibration
of the metallicities and ages (Haywood 2006; Holmberg et al. 2007;
Haywood 2008, Casagrande et al. 2011). 
Recently, the re-calibrated data from Holmberg
et al. (2009) became available, and it is with the MDF drawn from these data that we 
compare our model predictions. 
Following Holmberg et al. (2009), we define
a 'clean' sub-sample by removing (i) binary stars, (ii)
stars for which the uncertainty in age is $>$25\%, (iii) stars
for which the uncertainty in trigonometric parallax is $>13$\%,
and (iv) stars for which a 'null' entry was provided for any of
the parallax, age, metallicity, or their associated uncertainties.
This 'clean' sub-sample consists of  $\sim4000$ stars. 

The region including the star particles with  distances $r$ in the range 
2$R_{d}$$<$$r$$<$3$R_{d}$ are analogous `solar 
neighbourhoods' for the MUGS simulations;  in the following, 
we will employ the MDF calculated for the star particles included in these region for our 
comparison with extant data of the Milky Way's solar neighbourhood.  

The solid lines in Fig. ~\ref{mdf_sn} are the MDFs calculated after a kinematic 
cut as described in Sect.~\ref{kcut} plus a spatial cut 
as explained in Sect.~\ref{spcut}.  
The dashed lines are MDFs after applying a spatial cut and by excluding all the star particles 
with $J_{z}/J_{circ}(E) < 0.8$. 
A comparison of the solid and dashed lines helps in understanding the role of 
bulge stars with positive circularities in the 'solar neighbourhood' MDF:  
in all the panels, the MDFs computed in these two different ways are very similar, 
hence the contribution from bulge star particles at radial distances $> 2 R_{eff}$ is expected to be low. 
Also the MDFs computed by means of a solely spatial cut (dotted lines in Fig. ~\ref{mdf_sn}) are very similar to 
the ones computed with a spatial plus kinematic cut. 
This shows that our results concerning the ``solar neighbourhood'' MDF are
not overly sensitive to a kinematic selection of disc stars. 
This is an encouraging aspect, since for the observational sample 
we are comparing our results with, it is impossible to perform any such distinction between bulge and disc stars. 
The fact that the results are stable against the definition of ``solar neighbrouhood'' indicates that 
our results should be considered robust. 

Concerning the comparison between model results and the observed MDF (solid histograms in Fig. ~\ref{mdf_sn}),  
the first striking difference regards the position of the peak metallicities, with the model MDFs 
peaking at [Fe/H] values $\sim0.2 -0.3$ dex lower than the one of the GCS sample. 
This is an indication that the average stellar metallicity in the simulations 
is lower with respect to the one observed in the solar neighbourhood. 
Another important aspect emerging from this comparison regards the relative fraction of low metallicity stars, 
which in the simulated galaxies is substantially higher, and which will be quantified and discussed in more detail 
in Sect.~\ref{cumu}.

\subsubsection{Statistical analysis of the MDF in simulations and in the solar neighbourhood}
A comparison of the main statistical features of the predicted MDFs and 
the empirical MDF of the Milky Way's solar neighbourhood is provided in 
Tab.~\ref{mdf_feat}. 
The characteristics noted there are patterned 
closely upon those performed by Kirby et~al. (2011) in their analysis of 
the MDFs of Local Group dwarf spheroidals 
and by Pilkington et al (2012b) in their analysis of the MDFs of simulated dwarf disc galaxies. 
In the second and third 
columns, we list the mean and the median of the MDF for a given 
simulation, whose name is reported in the first column. 
In this table, 
we use 5-$\sigma$ clipping of outliers from the distribution before deriving MDF shape characteristics.
The dispersion 
$\sigma$ is reported in the fourth column, whereas the interquartile 
range (IQR), the interdecile (IDR), intercentile range (ICR), inter tenth percent (ITR), 
the skewness, and the kurtosis are reported in the fifth, 
sixth, seventh, eigth, ninth amd tenth columns, respectively. 
The dispersion $\sigma$, the  
IQR, IDR, ICR and ITR are different measures of the width of the distribution, while 
the skewness is a measure of the symmetry, with positive and negative 
values indicating an MDF skewed to the right (towards positive 
metallicities) and to the left (towards negative metallicities), 
respectively. The kurtosis (or peakedness) indicates the degree to which the MDF is 
peaked with respect to a normal distribution: 
high kurtosis values ($\gg 1$) signify a distribution with extended tails, while lower values signify light tails. 

As already noted, the 
model MDFs show median metallicities that are systematically lower than 
the empirical MDF, with offsets ranging from $-$0.14~dex (considering 
the median as representative of the peak position) to $-$0.45~dex. In the 
fiducial simulation (\tt g15784\rm), the offset between the peaks is 
$-$0.32~dex. 

The discrepancy is lower for \tt g24334\rm, for which 
we are only considering the innermost regions since its disk scalelength is only 1 kpc. 
It is 
therefore not surprising that the mean and median stellar metallicities 
for this system are larger with respect to the others, due to the 
presence of a significant metallicity gradient (Pilkington et~al. 2012a), 
a point to which we return below.

\subsubsection{Possible reasons for the discrepancies between simulations and observations}
The apparent discrepancy between the MDF peaks of the simulated 'solar 
neighbourhood' and that observed in the Milky Way can be ascribed to several  
reasons.  First, the chemical evolution prescriptions 
incorporated within \textsc{Gasoline} (Raiteri et~al. 1996) predict the 
cumulative Fe mass produced after 10 Gyr by a 1 $M_{\odot}$ simple 
stellar population, employing a Kroupa et~al. (1993) IMF is 
0.0007~M$_{\odot}$, a factor of two lower than that reported by 
Portinari et~al. (2004), in their study of the impact of the IMF on 
various local chemical evolution observables. This discrepancy is due in 
part to a different SNeIa frequency (7\% in the case of Portinari et al. 
2004 versus 5\% adopted within \textsc{Gasoline}), and to a lower Fe 
yield from SNeII (in \textsc{Gasoline}, SNeII form a $1M_{\odot}$ SSP 
produces 0.00022~M$_{\odot}$ of Fe, versus 0.00048~M$_{\odot}$ according 
to Portinari et al. 2004) and to a slightly lower mass of Fe produced by 
a single SNIa (0.63~M$_{\odot}$ versus 0.7~M$_{\odot}$ used by Portinari 
et~al. 2004). Therefore, the nucleosynthesis prescriptions adopted 
within \textsc{Gasoline} tend to produce less Fe with respect to other 
chemical evolution models designed to reproduce local constraints, and 
this is certainly one of the reasons for the lower metallicity peaks of 
the simulated MDFs. Another reason is linked to the formalism used here 
to model SNeIa: here, the contribution of SNeIa progenitors with mass 
$m$$<$1.5~M$_\odot$ was neglected, and this leads to the 
underestimation of the Fe mass in stellar particles older than 5 Gyr. 
In fact, by integrating the type Ia SN rate of Greggio \& Renzini (1983) 
from $T_{0}$=0 Gyr to $T_{1}$=5 Gyr, corresponding to the lifetime of a 
1.5~M$_\odot$ star, and comparing this number to the integral 
of the rate over one Hubble time, one can show that 
the total Fe production from stars down to the present turnoff mass (0.8~M$_{\odot}$) 
is underestimated by $\sim$0.1~dex. 

Other effects 
which could cause a loss of metals and a consequent low stellar 
metallicity in the disc  would be metal ejection during 
mergers occurring at early times. 
In this case, we should be finding mean metallicities higher in the 
gas (including both cold and warm components) with respect to the stars. 
However, a preliminary estimate of gas and stellar metallicity gradients in 
the simulated discs do not seem to support this scenario:  
in the g15784 simulation, 
the mass-weighted average metallicity calculated for 
the gas particles belonging to the most massive galaxy (i.e. the g15784 disc) 
is $<Z>_{g}=0.0005$, whereas the analogous stellar mean metallicity 
is $<Z>_{*}=0.007$.

Finally, as indicated by a parallel 
study of the evolution of the metallicity gradients in the MUGS galaxies 
(Pilkington et~al. 2012a), the star formation threshold may 
contribute to a more centrally concentrated star formation history, in 
particular during the early stages.
If star formation in the disk is underestimated in the models at early times,
this may lead to 
steep abundance 
gradients and low metallicity star formation in the early disc, resulting in a lower 
present-day metallicity in this region.

The model MDFs appear broader than the observed one, as indicated by the $\sigma$, IDR-ITR 
values. 
This could be related to radial migration 
of stellar particles, which can broaden the MDF (Schoenrich \& Binney 2009) and 
whose effect could be enhanced by the fact that 
these simulations (like most cosmological disc simulations) are 
substantially hotter (kinematically speaking) relative to the Milky Way 
(e.g. House et~al. 2011).  
Metal circulation in the disc and outwards could play some role as well in broadening the MDF.  

The skewness values vary from disc to disc, however, in most cases, the 
simulated MDFs tend to show more negative skewness with respect to the 
observed MDF of the solar neighbourhood, which relates mostly to their 
over-populated low-metallicity tails.  

Finally, the kurtosis values are higher than the one derived for the GCS sample, 
again due to heavy low metallicity tails. 
It is worth noting that the use of a 5-$\sigma$ clipping limits the effects of the 
presence of extreme low-metallicity tails which, without taking into account any clipping, 
would give rise to even higher kurtosis values. 

Later in Section~\ref{AMR}, we will see how the age-metallicity relation may be regarded as a useful diagnostic to 
understand in better detail 
the implications of our MDF study and the reasons for the discrepancies between the theoretical and obserevd MDFs. 


\begin{figure*}
\epsfig{file=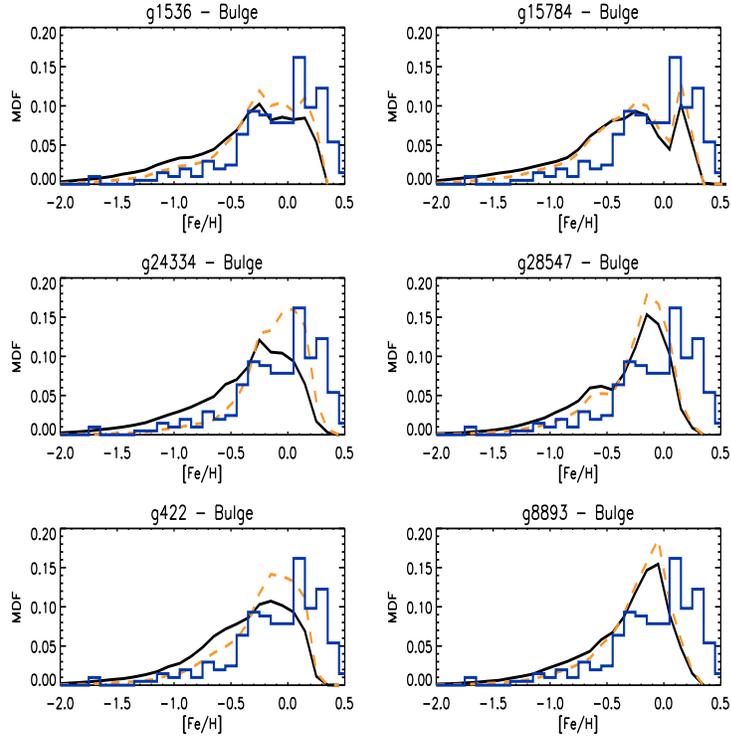,height=10cm,width=10cm}
\caption{Metallicity distribution functions for the bulges of each of 
the six MUGS galaxies. Solid lines: MDFs calculated by using all the 
particles belonging to the spheroid, after application of the kinematic 
decomposition described in \S\ref{kcut}.
Dashed lines: spheroid MDF
restricted to the innermost regions of the bulge, i.e. computed by 
performing a spatial cut on the bulge component. 
In each panel, the solid histogram is the observational bulge MDF from Zoccali et al. (2008). }
\label{mdf_bul}
\end{figure*}
\begin{figure*}
\epsfig{file=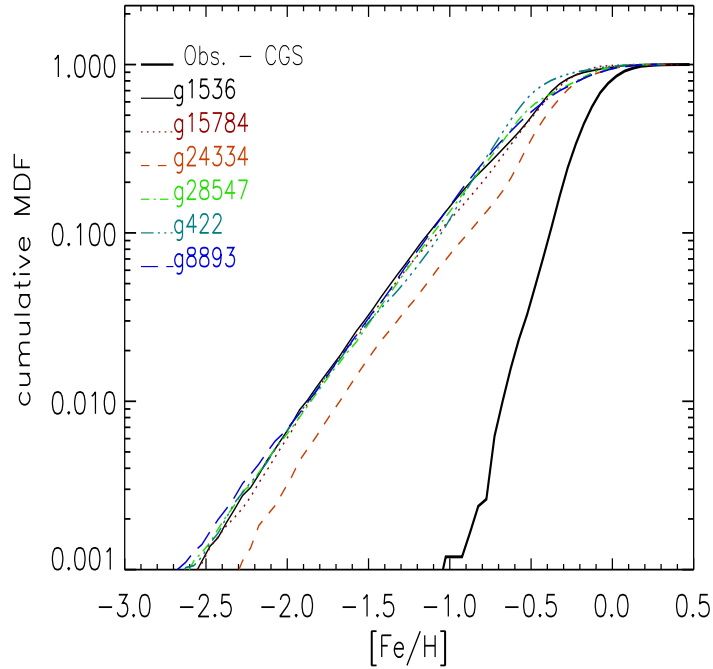,height=10cm,width=10cm}
\caption{Cumulative metallicity distribution functions for the 'solar 
neighbourhood' of each of the six MUGS galaxies and for 
the CGS 'clean' sub-sample, each one 
normalized at the integral of the corresponding differential MDF. }
\label{mdf_cumu_sn}
\end{figure*}
\begin{figure*}
\epsfig{file=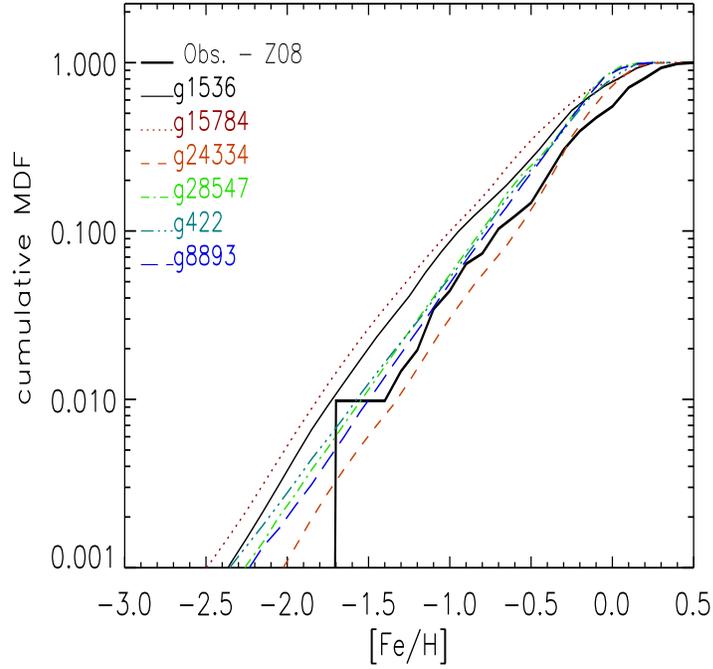,height=10cm,width=10cm}
\caption{Cumulative metallicity distribution functions for the bulges of 
each of the six MUGS galaxies and for the Z08 sample. 
Each cumulative MDF is normalized to the integral of the corresponding differential MDF. }
\label{mdf_cumu_bul}
\end{figure*}
\begin{figure*}
\epsfig{file=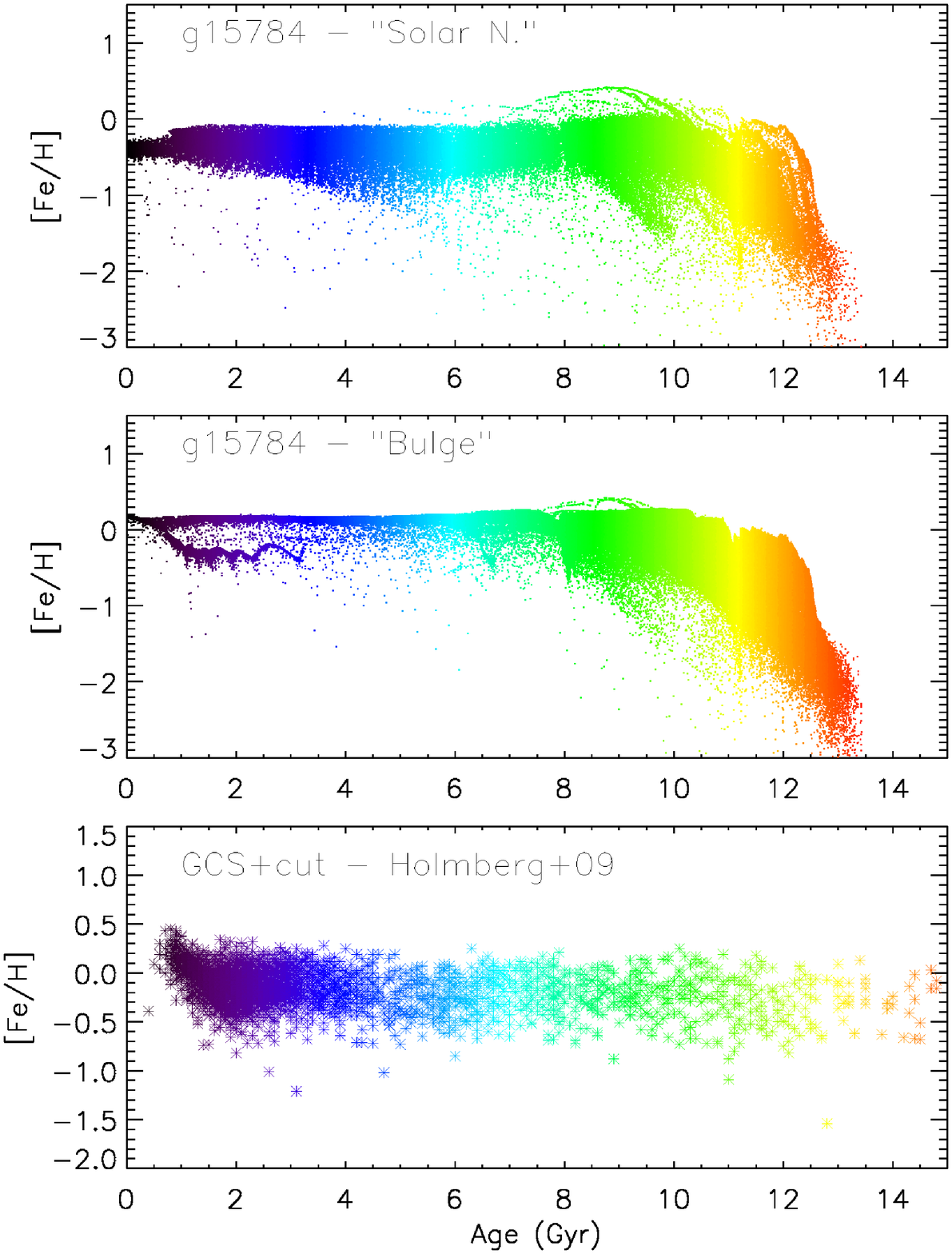,height=10cm,width=10cm}
\caption{Age-metallicity relations for the ``solar neighbourhood'' (upper panel) 
and of the ``bulge'' (middle panel) of the simulated galaxy g15784, 
and of the sub-sample of Holmberg et al. (2009) of solar neighbourhood stars 
described in Sec.~\ref{SN}. The colours of each simulated or observed star 
scales with its age (colour bar on the right).}
\label{agez}
\end{figure*}

\subsubsection{Possible selection effects in the observed sample}

In principle, two possible kinds of bias may affect 
the comparison between data
and simulations. One is connected with the observational uncertainties
and the sample selection, the other involves the "representativeness" of
the local sample. Concerning the former, the 'clean' sample of Holmberg et al. (2009) 
is complete down to $M_{V} \sim4.5$ up to about 40-50 pc, thus no
significant fraction of F-G stars is expected to be missed. In
principle, the choice of using F-G stars (which are long-lived enough to
trace the disk star formation over the entire Hubble time) may imply a
slightly different range of masses for populations of different
metallicities (a metal poor star is hotter than a metal rich star, so
for a fixed spectral type it will be slightly less massive), but this
effect is likely to be very small and it is unlikely to substantially affect our analysis. 

The other potential issue stems from the definition of "solar
neighborhood" itself. Dynamical diffusion of orbits\footnote{Stellar
velocities are randomized through chance encounters with interstellar
clouds, gaining energy and increasing the velocity dispersion (e.g.
Wielen 1977).} allows stars to drift from their birthplaces over time
scales of several Gyr and, as a consequence, may deplete the local (i.e. regarding the volume within $\sim100$ pc)
star formation rate at early epochs (see e.g. Schr\"oder \& Pagel 
2003). Indeed, different thin disk populations are known to have
different scale heights, with their height increasing with age. Since
oldest stars are likely the most metal poor, this could imply an
observed local metallicity distribution biased against lowest
metallicity stars. 

It is not possible to assess quantitatively the role of each of the effects described 
in this section. Important constraints may come from the Gaia mission, which will soon take a complete census of
stars down to ${M}_{V} = 4.5$ with parallaxes measured with an accuracy
better than 10\% up to 2–3 kpc, thus extending the solar neighborhood
sample to a realistic disk/thick disk sample. 

\subsection{Metallicity Distribution Functions in the Bulge}

The bulge MDFs for the six MUGS galaxies are shown in Fig~\ref{mdf_bul}, 
together with observational data from Zoccali et~al. (2008, Z08 hereinafter). 
The latter 
are derived from a survey of 800 K-giants in the Galactic bulge, 
observed at a resolution $R$$\sim$20000. In each panel, we show two MDFs 
derived from the simulations: one based upon the use of all star 
particles belonging to the bulge, after application of the kinematic 
decomposition described in \S\ref{kcut} (solid lines), and a second 
restricted to the innermost regions of the bulge, i.e. computed by 
performing a spatial cut on the bulge component (dashed lines). As 
described in \S\ref{spcut}, the stellar particles belonging to the bulge 
region are those at a distance $r = \sqrt{X^2+Y^2+Z^2}$ $<$ $R_{b,cut}$ 
(where, 1$<$$R_{b,cut}$$<$3~kpc).

The main features of the observed MDF for the Galactic bulge and 
those derived from the simulations are summarised in 
Table~\ref{bul_feat}. The entries reported in the columns of Table~\ref{bul_feat} 
are the same as those of Tab. ~\ref{mdf_feat}, shown in the same order. 
Both the empirical and simulation-based MDFs show 
negative skewness, and are more asymmetric than the corresponding MDFs 
for the solar neighbourhood. The empirical MDF of Zoccali et~al. (2008) 
shows a peak centered nearly at solar [Fe/H] and broader than the one of 
the solar neighbourhood. The bulge MDF also possesses a  
negative skewness as the solar neighbourhood. 
The higher kurtosis of the bulge MDF indicates a more peaked metallicity 
distribution with respect to that of the solar neighbourhood.

In most of the cases, the peak of the model MDF, represented by the mean 
and median values reported in Tab.~\ref{bul_feat}, is offset with 
respect to the empirical one by $-$0.2 - $-$0.3~dex; such an offset is 
perhaps not surprising, considering the implementation of the stellar 
yields, as discussed in \S\ref{SN}. The discrepancy is lower only for 
\tt g24334\rm, a simulation for which the stars accreted from disrupted satellites 
dominate over the ones formed in its main disc    
and one whose radial abundance 
gradient is both steep and shows little temporal evolution, while the 
others (e.g. \tt g15784 \rm and \tt g422\rm) show gradients which 
flatten with time (Pilkington et~al. 2012a).  Only \tt\,g24334\,\rm    
shows the same steep radial gradient today as it showed at redshift 
$z$$\sim$2.5, leading in part to higher overall metallicity in its 
innermost regions, which are the ones we are considering as ``solar neighbourhood'' owing to 
its small disc size. 
In all these senses, \tt g24334 \rm differs from the 
other five MUGS simulations, which each show similar mean and median 
[Fe/H] values. 
The dispersions of the simulated bulge MDFs 
are, in general, in better agreement with the observations than the case 
of the solar neighbourhood MDFs. A good agreement between model predictions 
and observations is visible also for the skewness values. 
The kurtosis values are still higher than the one of the 
observed bulge MDF, however the agreement is better than 
in the solar neighbourhood.

Two recent studies of red giants in Baade's Window (Hill et~al. 2011) 
and dwarf/subgiants in the Galactic bulge (Bensby et~al. 2011) suggest 
that the MDF in the inner region has multiple peaks (or is at least 
double-peaked), with a low-metallicity peak occuring near 
[Fe/H]$\sim$$-$0.6 (Bensby et~al.) or [Fe/H]$\sim$$-$0.3 (Hill et~al.), 
and a higher-metallicity peak centered at [Fe/H]$\sim$$+$0.3. Hill 
et~al. suggest that the low-metallicity stellar component shows a large 
dispersion, while the high-metallicity component appears narrow.
 
The above results are in qualitative agreement with those shown in 
Fig~\ref{mdf_bul}, in particular as far as the Milky-Way analogue \tt 
g15784 \rm is concerned. A low-metallicity component peaked at 
[Fe/H]$\sim$$-$0.3 showing significant dispersion is visible in the 
top-right panel of Fig~\ref{mdf_bul}, with a narrower component centered 
near [Fe/H]$\sim$$+$0.15. The relative amplitude of the two peaks in \tt 
g15784 \rm is close to unity, while the aforementioned studies of Hill 
et~al. (2011) and Bensby et~al. (2011) suggest that the 
lower-metallicity peak appears considerably weaker than the one at 
high-metallicity. Both studies converge to a picture of a longer 
formation time-scale for the metal-rich component. Such a picture is 
consistent with the age-metallicity relation predicted for the bulge of \tt g15784\rm, 
(see Sect.~\ref{AMR}), and with the star formation histories shown in Fig~1. 

From Fig~1, it is clear that the peak of the star formation occurs at 
early times.  Additionally, while residual star formation is present at 
relatively recent times, this does not contribute to the bulk of the 
stellar mass, i.e. most of the stars in the simulated bulges are old and 
have low metallicity. 

Another aspect emerging from the studies of Hill et~al. (2011) and 
Bensby et~al. (2011) is the uncertainty in the position/metallicity of 
the metal-poor stellar component, likely due to the very different 
sample selection criteria between the two studies. For the moment, we do 
not attempt to perform any more detailed comparison with these results 
since they are very recent and awaiting further confirmation.

\subsection{The Cumulative Metallicity Distribution}
\label{cumu}
A diagnostic often used to investigate in better detail the low metallicity tail of the MDF 
in chemical evolution models is the cumulative metallicity distribution 
function. The cumulative MDF, calculated at a given metallicity [Fe/H], 
represents the number of stars with metallicity lower than [Fe/H]. 

The cumulative MDF reflects essentially the same information as the 
differential MDF (Caimmi 1997), but it is less sensitive to small number 
statistics and better tracks the behaviour at low metallicties for both 
low-metallicity local galaxies (Helmi et al. 2006) and the solar 
neighbourhood. In Fig.~\ref{mdf_cumu_sn}, we show the cumulative MDF as 
observed in the solar neighbourhood and as predicted for the six MUGS 
galaxies. 
In Fig. ~\ref{mdf_cumu_sn}, each curve is normalized to the integral of 
the corresponding MDF.

From the upper of Fig~\ref{mdf_cumu_sn}, one can see that in the 
Milky Way's solar neighbourhood, from the sample of Holmberg et al. (2009) 
and with the cuts performed in Sect.~\ref{SN}, $\sim$10\% of the stars have a 
metallicity [Fe/H]$<$$-$0.5. 
This is in stark contrast with the simulation results which, without taking into account 
colour and magnitude selection effects,  below [Fe/H]=-0.5 
show fractions greater than $\sim$30\%. 

Mass fractions relative to the total MDF for the stars associated with the four metallicity values 
are tabulated in Tab. ~\ref{tab_mdf_cumu_sn}, where the excess of low metallicity stars is 
emphasised: at any metallicity value from [Fe/H]=-2 to [Fe/H]=$-0.13$, 
the predicted mass fraction is considerably 
higher than the values obtained by integrating the observational MDF. 
This discrepancy  is certainly related to the metallicity offset visible 
in the peaks of differential MDF discussed in Sect.~\ref{SN}. 
However, the excess of very low metallicity stars is to be ascribed to other reasons. 
Several solutions to this problem could be plausible: 
one of them is the adoption of a modified IMF, known to alleviate the 
excess of metal-poor stars 
in local dwarf galaxies - (see e.g., Calura \& Menci 2009). 
However, since this would cause a larger relative number of high mass 
stars in a stellar population, this modified IMF would have a strong 
impact on the abundance ratios, such as the [O/Fe] ratio, and 
even on the SN feedback. In the 
solar neighbourhood, an IMF similar to that of Kroupa et~al. (1993), as 
adopted here, reproduces a large set of observational constraints, 
including the abundance ratios (Calura et al. 2010). Further, since a 
truncated or even slightly top-heavy IMF is known to produce a strong 
$\alpha$ enhancement in the abundance ratios (Calura \& Menci 2009), 
this does not seem to represent a proper solution to our problem. 
Investigations of the abundance ratios in the solar neighbourhood of the 
MUGS discs will be a useful to test this hypothesis as a solution to 
the  problem related to the excess of metal-poor stars. 
Such an analysis is currently underway, but 
deferred to a forthcoming paper.

An alternate explanation to the dearth of low metallicity stars in the 
solar neighbourhood is to invoke a prompt initial enrichment scenario, 
with a population of objects such as zero metallicity (Pop~III) stars 
(Matteucci \& Calura 2005; Ohkubo et al. 2006; Greif et al. 2007), 
releasing a sufficient amount of heavy elements to prevent the formation 
of very low-metallicity stars in galactic discs. Testing the impact of 
such objects in the chemical enrichment of discs in simulations would be 
possible by including the yields of very massive stars as provided by, 
e.g. Ohkubo et al. (2006). This is beyond the scope of the present 
paper, but under consideration for a future generation of 
chemo-dynamical simulations.

In Fig.~\ref{mdf_cumu_bul}, we show the cumulative predicted bulge MDF 
for our MUGS galaxies and that observed in the Galactic bulge by Zoccali 
et~al. (2008), with each cumulative MDF normalized to the integral of the corresponding 
differential MDF. 
The mass fractions relative to the total MDF for the stars associated with the various  metallicity values 
are tabulated in Tab. ~\ref{tab_mdf_cumu_bulge}. 
Also in the bulges  the simulated galaxies 
overestimate the number of low metallicity stars at most metallicities. 
However, aside from the overabundance of low-metallicity stars, 
the rest of the form of the model MDF is quite consistent with the observational MDF of Z08. 
Again, it will be interesting to see 
in the future if the predicted overabundance of low metallicity stars is 
due to sampling effects in the observational 
MDF and if this could be alleviated with different nucleosynthesis 
prescriptions for zero metallicity stars. 

\subsection{The Age-Metallicity Relation}
\label{AMR}
In Fig.~\ref{agez}, we are showing the age-metallicity relation (AMR) in the ``solar neighbourhood'' of our 
fiducial Milky Way analogue \tt g15784 \rm (upper panel), of its bulge (middle panel) and of our GCS 'clean' 
sub-sample (lower panel). From this figure, it is possible to appreciate how the slope of the AMR 
is reflecting the MDF problems described in the previous sections, 
in particular 
the over-populated low metallicity tails visible in the theoretical MDFs and their negative skewness 
(see also the parallel study of the MDF in dwarf spiral galaxies of Pilkington et al. 2012b).  
 
In Fig.~\ref{agez}, the metallicities of the stars in the Milky
Way’s solar neighbourhood (GCS) are higher than those of the \tt g15784 \rm ``solar neigbourhood''. 
This should not be surprising, in the light of the discussion in Sect.~\ref{SN}, in particular regarding 
the MDF peak metallicity of our simulations, lower than the one of the GCS. 

A striking feature of Fig.~\ref{agez} is that  
the AMR of the GCS solar neighbourhood sub-sample 
is essentially flat. 
The AMR predicted for the \tt g15784 \rm solar neighbouhood  is predominantly flat, 
but significant deviations are clearly 
visible at ages $> 10$ Gyr, where the 
\tt g15784 \rm simulation shows essentially a correlated AMR. 
In shape, this relation is similar to 
those predicted by classical galactic chemical evolution
models (e.g. Fenner \& Gibson 2003, Haywood 2006; Spitoni et al. 2009). 

These hints for a correlated AMR are 
to be associated to the excessive negative skewed MDFs of the simulations: 
the excessively negative skewness is mostly due to low metallicity 
star particles with ages $>10$ Gyr, older than the stars which settle on the plateau of  
AMR relation. 

Furthermore, the dearth of stars 
older than 13 Gyr is striking as well, and is at variance with the results of  
non-cosmological homogeneous galactic chemical evolution models (Matteucci et al. 2009), 
which in general are successful in reproducing the local MDF.  
Perhaps more hints on this aspect could come from studies with 
inhomogenous 
chemical evolution models (Oey 2000; Cescutti 2008) , which allow to 
explore the parameter space nearly as fast as homogeneous models, 
but more suited 
to explore the causes of asymmetric or multi-peaked MDFs or 
the dispersion and slope of the AMR.

Also the bulge of \tt g15784 \rm sees a flat age-metallicity relation for 
stars born with  [Fe/H]$\sim$$+$0.15. 
The formation of these stars,  
which would populate the high-metallicity peak in the bulge MDF, 
extends over a period of $\sim$10~Gyrs or more. Most of the stars 
older than $\sim$10~Gyrs have clearly formed with a lower metallicity 
and present a correlated AMR similar to the solar neighbourhood of \tt g15784 \rm. 
A more-thorough study of the AMR of MUGS simulations in both bulges and discs  
will be presented in Bailin et al., in prep.

\subsection{MDFs of In-situ Stars vs Accreted Stars}

Important clues as to the origin of various stellar populations formed 
at different metallicities may come from the study of the separate MDFs 
of the stars born \emph{in situ} as opposed to those \emph{accreted} 
(Sales et~al. 2009; Roskar et~al. 2008; Sanchez-Blazquez et~al. 2009; Rahimi et~al. 
2010). We define \it in situ \rm stars as being those born within a 
cylinder with a radius linearly increasing with cosmic time, constrained 
to a present-day current value of $\sim$30~kpc, with a time-independent 
height of $\pm$2~kpc (House et al. 2011). The choice of these values for radius and height 
were suggested after visual inspection of the appearence of the galactic 
disc at various cosmic times and redshifts. The choice for the height 
does not have any impact on our results (for values on the order of a 
few kpc); the value chosen is conservative and satisfies our desire to 
exclude the effects of satellites.  In other words, \it in situ \rm 
stars are born `locally', at small distances from the main progenitor, 
while \it accreted \rm stars form within satellites and are accreted 
later.  Such a cylindrical volume encompasses both bulge and disc stars; 
no kinematical distinction is performed in this case, hence our results 
are valid for the whole galaxy including all its kinematical components.

In this section, we will only show the 
results regarding our Milky Way fiducial (\tt g15784\rm) - partly
because it does represent the best analogue to our own 
Galaxy (see also Brook et~al. 2011; 2012), and partly because its
output temporal cadence was the highest, ensuring the greatest 
wealth of data with which to work.  We did examine all relevant 
metrics which could be derived with the more limited information
available to us at the time of analysis and find that none of our 
results are tied to this one system.  This is consistent with the
results of Pilkington et~al. (2012a), regarding the similarity
of the metallicity gradients within the MUGS galaxies.\footnote{Modulo
the aforementioned discrepant (but understood) \tt g24334\rm.}

\begin{figure*}
\epsfig{file=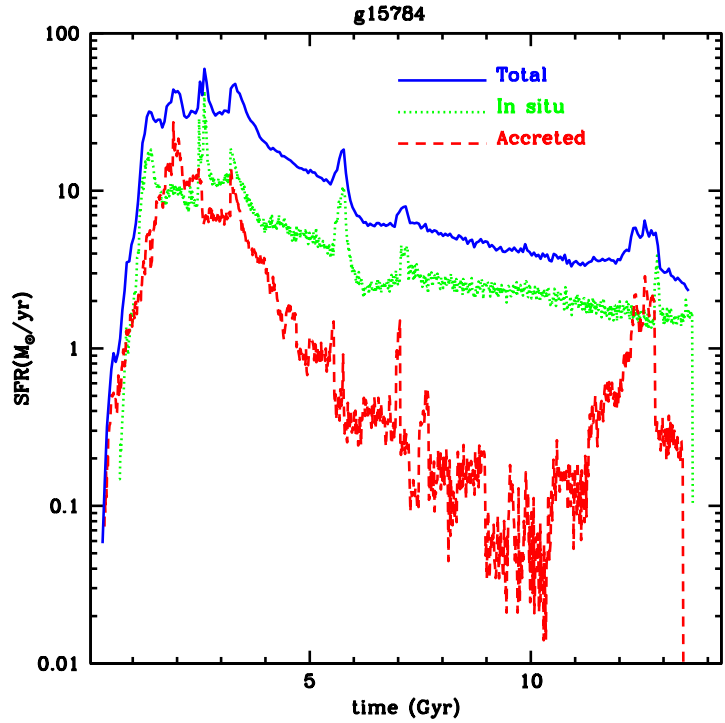,height=10cm,width=10cm}
\caption{\it In situ \rm (dotted line) vs \it accreted \rm (dashed line) 
star formation history  as a function of cosmic time for the MUGS galaxy g15784. 
The solid line 
corresponds to the the total (\it in situ \rm + \it accreted\rm) star 
formation history.}
\label{sfr_ia}
\end{figure*}

The SFHs of the stellar particles classified as \it in situ \rm and \it 
accreted \rm are shown for the MUGS galaxy \tt g15784 \rm in Fig.~\ref{sfr_ia}. The accreted stars form 
mostly at early times ($t_{\rm form}$$<$2~Gyrs, where the maximum of the 
`accreted' SFH lies). Another peak within the `accreted' SFH occurs more 
recently (12$\le t_{\rm form}\le$13~Gyr). The accreted stars 
contribute roughly 1/3 of the present-day stellar mass of the system; 
the majority of the stars are born \it in situ \rm via a merger event 
and/or local star formation episode.

\begin{figure*}
\epsfig{file=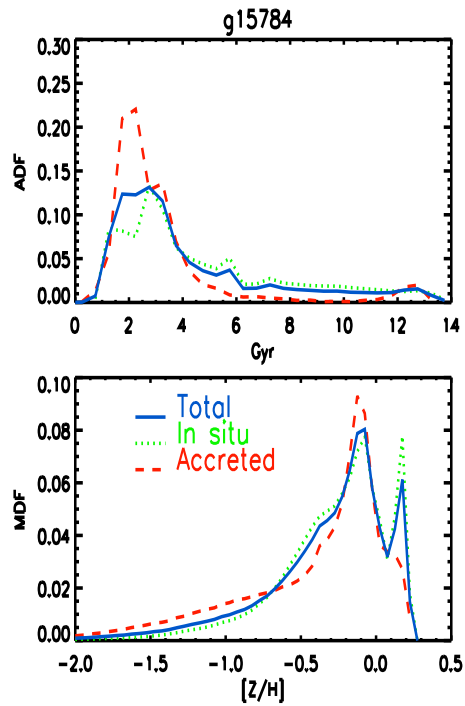,height=10cm,width=10cm}
\caption{\it In situ \rm (dotted line) vs \it accreted \rm (dashed line) 
metallicity distribution functions (lower panel) and age distribution 
functions (upper panel) for the MUGS galaxy g15784. The solid lines correspond to the total (\it in 
situ \rm + \it accreted\rm) distribution functions.}
\label{mdf_ia}
\end{figure*}

In Fig.~\ref{mdf_ia}, we show the total MDF (solid lines in both panels, 
normalised to unity in all cases) for \tt g15784\rm, alongside the MDFs 
computed considering only the stars born \it in situ \rm (dotted lines) 
and those \it accreted \rm (dashed lines). We show here the MDFs based 
upon the total global metallicity Z, rather than just Fe, as the primary 
MDF metrics in which we are interested are insensitive to this choice.

The \it in situ \rm MDF does not differ significantly from that of the 
total : both functions show two peaks (at [Z/H]=$-$0.2 and 
[Z/H]=$+$0.1). While we have not performed any kinematic decomposition 
here, it is quite obvious from the results discussed previously that the 
high metallicity peak is populated by the stars of the central region 
(i.e. bulge), whereas the lower-metallicity peak is generated by disc 
stars.  The \it accreted \rm stellar populations result in an MDF 
centered at [Z/H]$\sim$$-$0.1; furthermore, 
the MDF for accreted stars does not show any peak at higher metallicities present in the 
in situ MDF, and the 
low metallicity tail is 
more pronounced than that seen within the \it in situ \rm population, 
indicating a larger relative fraction of low metallicity stars. 
It is not surprising 
that the accreted stars have lower metallicity, as they come from disrupted satellites 
which have lower mass than the central galaxy 
\tt g15784\rm, hence they should have lower metallicity stars, 
consistent with what the well-known mass-metallicity relation suggests 
in local and high-redshift galaxies (Maiolino et al. 2008; Calura et al. 
2009).

\begin{figure*}
\epsfig{file=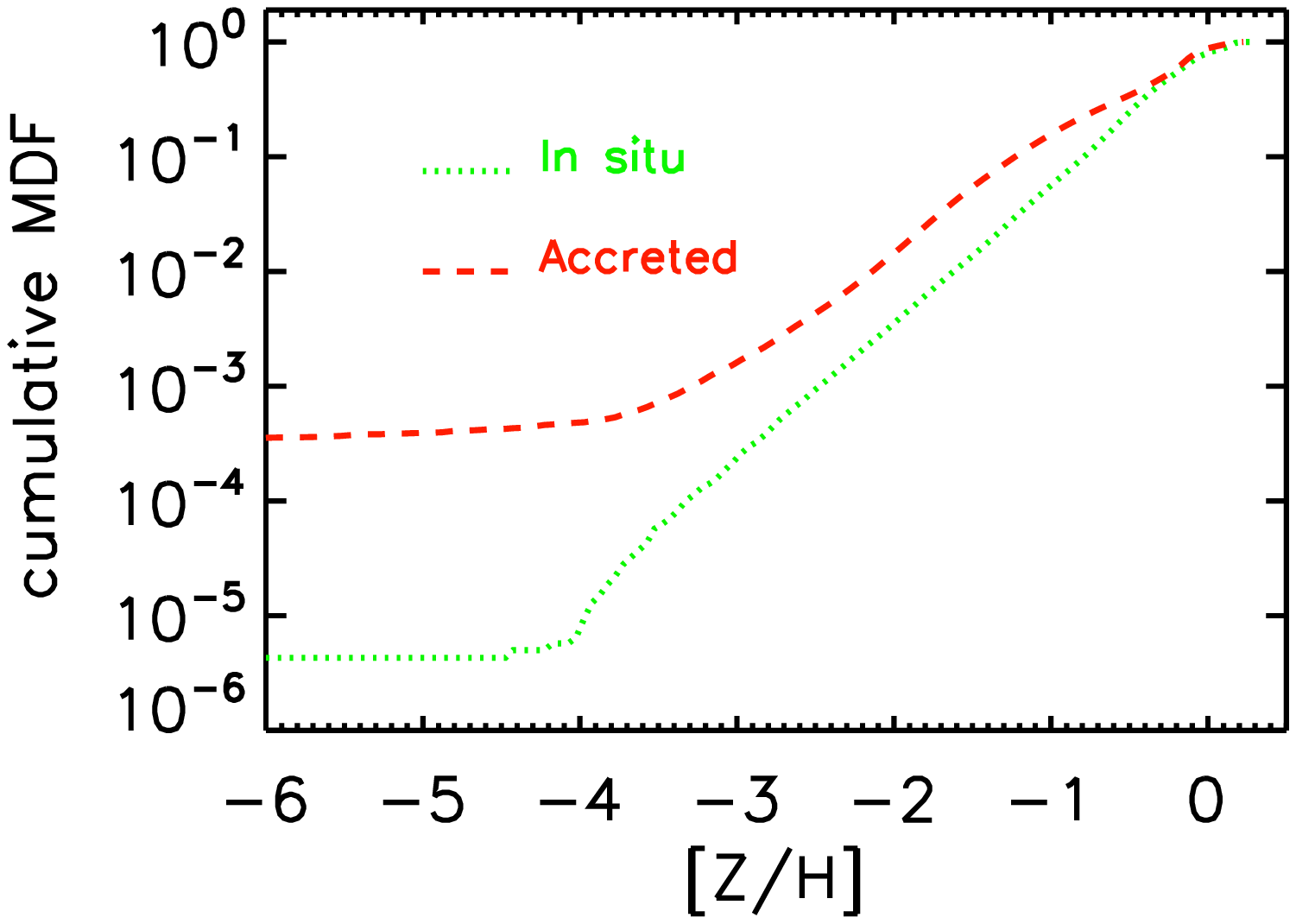,height=10cm,width=10cm}
\caption{\it In situ \rm (dotted line) vs \it accreted \rm (dashed line) 
cumulative metallicity distribution functions for the MUGS galaxy g15784. The two curves are 
normalised to unity at [Z/H]=$+$0.5.}
\label{mdf_cumu_ia}
\end{figure*}

In the upper panel of Fig.~\ref{mdf_ia}, we show the age 
distribution function (ADF) of the stars born \it in situ \rm and those 
\it accreted\rm.  The ADF shows that the accreted stars are mainly older 
than 10~Gyrs, whereas the stars formed in situ show a broad range of 
ages. \\
It is interesting to examine the cumulative MDF calculated for the stars 
born \it in situ \rm and for those \it accreted\rm ~(Fig.~\ref{mdf_cumu_ia}). The low metallicity 
tail (i.e. stars with [Z/H]$<$$-$4) is populated almost exclusively by 
accreted stars: this means that for \tt g15784\rm, the lowest 
metallicity stars are of extragalactic origin. 
We have investigated the 
kinetic energy of the stellar particles belonging to the two 
populations, but no clear distinction was found. This is consistent with 
that found by Rahimi et~al. (2010), i.e. that an early accretion event 
is unlikely to leave any strong signature in any obvious physical 
property of the present-time stellar populations. 

\begin{table*}
\vspace{0cm}
\begin{flushleft}
\caption[]{Main features of the MDF observed in the Milky Way's solar 
neighbourhood based upon the
Holmberg et al. (2009) GCS empirical dataset and those of the simulated ``solar neighbourhood'' 
MDFs for the six MUGS galaxies.}
\begin{tabular}{l|llllllllr}
\noalign{\smallskip}
\hline
\hline
\noalign{\smallskip}
                    &   Mean           &  Median         & $\sigma$       & IQR     & IDR    & ICR      & ITR    & Skewness   &   Kurtosis   \\
                    &   [Fe/H]         &  [Fe/H]         &                &         &        &          &        &            &              \\
\hline                                                                                                                                                       
Obs. MDF            &   -0.14         &  -0.12          &  0.18          &  0.23   &  0.44    & 0.92    & 1.42   &  -0.37      & 0.77    \\
(GCS)               &                 &                 &                &         &          &         &        &             &         \\
\noalign{\smallskip}                                                                                                                                         
\hline
\tt g1536  \rm              &   -0.56         &   -0.47          &  0.40           & 0.44    &  0.96   & 1.97    & 2.76   &  -1.12     &  1.82   \\ 
\tt g15784 \rm             &   -0.54         &   -0.46         &  0.37           & 0.40    &  0.88   & 1.89    & 2.90   &  -1.31      &  2.37  \\
\tt g24334 \rm             &   -0.43         &   -0.38         &  0.34           & 0.33   &   0.80   & 1.78    & 2.56    & -1.29      &  2.74  \\ 
\tt g28547 \rm             &   -0.56         &   -0.53         &  0.40           & 0.51   &   0.99   & 1.91    & 2.71   & -0.93      &  1.40  \\
\tt g422   \rm              &   -0.62         &   -0.58         &  0.36           & 0.37   &   0.82   & 1.97    & 2.84   &  -0.91     &  2.35   \\
\tt g8893  \rm              &   -0.55         &   -0.50         &  0.42           & 0.55   &   1.05   & 2.03    & 2.88   &  -0.85     &  1.19   \\  
\hline
\hline
\end{tabular}
\label{mdf_feat}
\end{flushleft}
\end{table*}

\begin{table*}
\vspace{0cm}
\begin{flushleft}
\caption[]{Main features of the MDF observed in the Milky Way's bulge 
and those of the simlated ``bulge'' MDFs for the six MUGS galaxies.}
\begin{tabular}{l|llllllllr}
\noalign{\smallskip}
\hline
\hline
\noalign{\smallskip}
                    &   Mean           &  Median         & $\sigma$       & IQR     & IDR    & ICR      & ITR    & Skewness   &   Kurtosis   \\
                    &   [Fe/H]         &  [Fe/H]         &                &         &        &          &        &            &              \\
\hline                                                                                                                                                       
Obs. MDF            &   -0.05         &  0.05            &  0.40          & 0.51     & 0.95   &  1.74    & 2.24    &  -1.23     & 1.89        \\
(Z08,Baade's window)&                 &                  &                &         &        &          &         &             &                \\
\noalign{\smallskip}                                                                                                                                         
\hline
\tt g1536    \rm            &   -0.29         &   -0.22          &  0.42          & 0.51    &  1.06  & 1.92   & 2.58      &  -1.23    &  1.75             \\ 
\tt g15784   \rm            &   -0.35         &   -0.28          &  0.45          & 0.58    &  1.11  & 2.02   &  2.72     &  -1.05    &  1.36               \\
\tt g24334   \rm            &   -0.16         &   -0.10          &  0.30           & 0.35   &  0.68  & 1.57   &  2.30     &  -1.42   &   2.80               \\ 
\tt g28547   \rm            &   -0.29         &   -0.19          &  0.34           & 0.39   &  0.81  & 1.70   & 2.48      &  -1.40   &   2.33               \\
\tt g422     \rm            &   -0.26         &   -0.17          &  0.36           & 0.43   &  0.86  & 1.74   & 2.55      &  -1.37   &   2.30      \\
\tt g8893    \rm            &   -0.27         &   -0.18          &  0.33           & 0.38   & 0.78   & 1.66   &  2.41     &  -1.40   &  2.45             \\  
\hline
\hline
\end{tabular}
\label{bul_feat}
\end{flushleft}
\end{table*}

\begin{table*}
\vspace{0cm}
\begin{flushleft}
\caption[]{Cumulative MDF calculated at various metallicity values for the observed solar neighbourhood and for the solar neighbourhood analog sample 
of our six MUGS galaxies.}
\begin{tabular}{l|lllllll}
\noalign{\smallskip}
\hline 
\hline 
  [Fe/H]            &   Obs.    &  \tt g1536 \rm  & \tt g15784 \rm   & \tt g24334 \rm  &  \tt  g28547 \rm   &  \tt g422 \rm  &  \tt    g8893 \rm\\
\hline      
-2.000   &  0.000   &  0.006  &   0.006  &   0.003   &  0.006   &  0.006   &  0.006     \\    
-0.660   &  0.010   &  0.314  &   0.287  &   0.177   &  0.349   &  0.396   &  0.357     \\    
-0.450   &  0.048   &  0.524  &   0.513  &   0.395   &  0.587   &  0.701   &  0.547      \\   
-0.315   &  0.164   &  0.766  &   0.742  &   0.629   &  0.712   &  0.843   &  0.698      \\   
-0.130   &  0.446   &  0.898  &   0.915  &   0.831   &  0.843   &  0.920   &  0.830      \\   
\hline
\hline
\end{tabular}
\label{tab_mdf_cumu_sn}
\end{flushleft}
\end{table*}
\begin{table*}
\vspace{0cm}
\begin{flushleft}
\caption[]{Cumulative MDF calculated at various metallicity values for the observed Bulge  and for the ``bulge'' of our six MUGS galaxies.}
\begin{tabular}{l|lllllll}
\noalign{\smallskip}
\hline 
\hline 
  [Fe/H]            &   Obs.    &   g1536    &     g15784   &  g24334    &    g28547    &       g422      &      g8893 \\
\hline                    
-2.000  &   0.000   &  0.003  &   0.004   &  0.001   &  0.002   &  0.002   &  0.002     \\    
-1.230  &   0.015   &  0.041  &   0.052   &  0.013   &  0.025   &  0.025   &  0.022     \\    
-0.820  &   0.064   &  0.123  &   0.140   &  0.048   &  0.088   &  0.083   &  0.078      \\   
-0.351  &   0.211   &  0.304  &   0.388   &  0.159   &  0.271   &  0.263   &  0.255      \\   
 0.050  &   0.549   &  0.817  &   0.807   &  0.813   &  0.947   &  0.873   &  0.923      \\   
\hline
\hline
\end{tabular}
\label{tab_mdf_cumu_bulge}
\end{flushleft}
\end{table*}

\section{Conclusions}

We have analysed the MDFs constructed from a suite of six 
high-resolution hydrodynamical disc galaxy simulations. Both kinematic 
decompositions and spatial cuts were performed on each, in order to isolate 
samples of analogous `solar neighbourhood' and `bulge' samples, for 
comparison with corresponding datasets from the Milky Way.  Our main 
conclusions can be summarised as follows.

\begin{itemize}
\item In general, after having performed a kinematical decomposition of 
discs and bulges, in most of the cases the star formation histories of the 
discs dominate over those of the bulges. 
On the other hand, if we define discs and bulges on a solely spatial basis, 
bulges have unnaturally high present-day 
SFR values , which reflect the well-known issue of 
an excessive central concentration of 
mass in the simulated galaxies. 
Increasing resolution may help alleviate this problem, 
however it is not yet clear to what extent. 
\item At the present time, 
an excess of star formation is visible in the simulated bulges. 
To limit this phenomenon, 
the next generation of simulations will have to include 
mechanisms of star formation quenching such as AGN feedback, which, 
in semi-analytic models of galaxy formation, have turned out to be 
efficient in decreasing star formation timescales in spheroids.
\item The `oscillating' behaviour 
of star formation in the outermost parts is due to the adopted star 
formation density threshold which acts to control the ability of the 
low-density regions in the outskirts to undergo substantial and 
sustained star formation.
\item The MDFs derived using all the stellar particles situated within 
the virial radius possess a number of `peaks', each associated with stellar 
populations belonging to various kinematic components. Applying a 
spatial cut, in particular 
removing central stars within $R_{b,cut}$=1$-$3~kpc from the centre, 
has a significant effect on the MDF in that the highest 
metallicity peak is generally removed. As in nature, these highest 
metallicity stellar particles tend to reside in the central region, on 
average.
\item A region analogous to the solar neighbourhood was defined in each 
system, by considering all the star particles contained within an 
annulus 2$R_{d}$$<$$r$$<$3$R_{d}$. The MDF of these stars show 
median metallicities lower by 0.2$-$0.3~dex than that of the Milky Way's 
solar neighbourhood.  This can be traced to several reasons, including 
the lower stellar yields implemented within \textsc{Gasoline}.
The predicted distributions are broader than the one observed in the solar 
neighbourhood, consistent with other 
studies of the MDF with cosmological simulations (Tissera et al. 2012). 
In our simulations, the overly broad MDFs are related to 
discs kinematically hotter relative to the Milky Way (House et al. 2011). 
The derived MDFs possess, on average, more negative skewness and higher kurtosis 
than those  
encountered in nature; such a result is traced to the more highly 
populated low-metallicity `tails' in the simulations' MDFs. 
\item The MDFs derived for stars in the bulges are in reasonable 
agreement with that observed in the Milky Way. While the median 
metallicities are somewhat lower, the inferred MDFs' dispersions and skewness are 
consistent with the Milky Way.  
The kurtosis values are higher than the one of the observed bulge MDF, 
however the agreement is better than in the solar neighbourhood. 
\item The prevalence of the low-metallicity tails are emphasised by 
examining the cumulative MDFs. In the solar neighbourhood, the predicted 
relative number of stars with [Fe/H]$<$$-$3 with respect to the number 
of stars with [Fe/H]$<$$-$2 is of the order 10\%, whereas in the Milky 
Way this ratio is effectively zero.  Solutions to this problem include 
prompt initial enrichment by a population of short-lived zero 
metallicity stars or perhaps an alternate treatment of metal 
diffusion (Pilkington et~al. 2012b).
\item For the fiducial simulation (\tt g15784\rm), we studied the star 
formation history and the MDF of the stellar populations born \it in 
situ \rm and of the those \it accreted \rm subsequent to satellite 
disruption. An early accretion episode generated a population of stars 
older than $\sim$10~Gyrs, whereas the stars formed \it in situ \rm show 
a broad range of ages. The low-metallicity tail of the MDF is populated 
mostly by accreted stars; this means that for \tt g15784\rm, the 
majority of the lowest metallicity stars are of extragalactic origin.
\end{itemize}
In the future, we plan to extend our study 
on the metallicity distribution function in MUGS galaxies to 
various kinematical components, including the halos, 
and to examine its variation with respect to position in the galaxy. 
A recent extensive work which will be useful to test 
our simulations is the one of Schlesinger et al. (2012), where 
by means of the SEGUE sample of G and K dwarf stars, 
the variations of the MDF in the Galaxy with radius and height 
has been investigated. 
A comparison of the results from various suites of disc galaxy simulations, 
such as Pilkington et al. (2012a), will be also useful to better understand  
the effects of the sub-grid physics on the MDFs.


\section*{Acknowledgments}
FC 
wish to thank Simona Bellavista for some useful suggestions and M. Bellazzini for 
interesting discussions. 
BKG, CBB and LM-D acknowledge the support of the UK's Science \& Technology 
Facilities Council (ST/F002432/1, ST/G003025/1). BKG, KP, and CGF 
acknowledge the generous visitor support provided by Saint Mary's 
University and Monash University. This work was made possible by the 
University of Central Lancashire's High Performance Computing Facility, 
the UK's National Cosmology Supercomputer (COSMOS), NASA's Advanced 
Supercomputing Division, TeraGrid, the Arctic Region Supercomputing 
Center, the University of Washington and the  
Shared Hierarchical Academic Research Computing Network (SHARCNET). 
This paper makes use of simulations performed as part of the SHARCNET Dedicated Resource project: 
'MUGS: The McMaster  Unbiased Galaxy Simulations Project' (DR316, DR401 and DR437).  
We thank the DEISA consortium, 
co-funded through EU FP6 project RI-031513 and the FP7 project 
RI-222919, for support within the DEISA Extreme Computing Imitative. 

\label{lastpage}
\end{document}